\documentclass[10pt,a4paper]{article}
\usepackage[english]{babel}
\usepackage[latin1]{inputenc}
\usepackage{amsfonts,amsbsy,bm,euscript,mathrsfs}
\usepackage{amssymb,stmaryrd,faktor}
\usepackage[tbtags]{amsmath}
\usepackage{bbm}
\usepackage{graphicx}
\usepackage[title,titletoc]{appendix}
\usepackage[bookmarks=true,colorlinks=true,linkcolor=blue,citecolor=blue,urlcolor=blue,bookmarksnumbered]{hyperref}

\usepackage{dsfont}
\usepackage{collref}
\usepackage{lmodern}
\usepackage{mathrsfs}
\usepackage{mathtools}
\usepackage{bbm}
\usepackage{braket}
\usepackage{slashed}

\usepackage{graphicx}
\usepackage{booktabs}
\usepackage{subfig}
\usepackage{tikz}
\usetikzlibrary{plotmarks,calc,decorations,decorations.pathmorphing}

\numberwithin{equation}{section}

\makeatletter
\renewcommand\section{\@startsection {section}{1}{\z@}
{-3.5ex \@plus -1ex \@minus -.2ex}
{2.3ex \@plus.2ex}
{\normalfont\Large\bfseries}}
\renewcommand\subsection{\@startsection{subsection}{2}{\z@}
{-3.25ex\@plus -1ex \@minus -.2ex}
{1.5ex \@plus.2ex}
{\normalfont\large\bfseries}}
\makeatother

\newcommand{\arXivlink}[1]{\href{http://arXiv.org/abs/#1}{arXiv:#1}}

\newcommand{\AdS}{\textup{AdS}}

\def\h{\eta}

\def\r{\rho}

\def\c{\chi}

\usepackage{blkarray}

\begin{document}

\thispagestyle{empty}
\begin{flushright}\footnotesize\ttfamily
DMUS-MP-20/09\\
TCDMATH-20-14
\end{flushright}
\vspace{2em}

\begin{center}

{\Large\bf \vspace{0.2cm}
{\color{black} \large Free Fermions, vertex Hamiltonians, and lower-dimensional $AdS/CFT$}} 
\vspace{1.5cm}

\textrm{Marius de Leeuw${}^*$\footnote{\texttt{mdeleeuw@maths.tcd.ie}}, Chiara Paletta${}^*$\footnote{\texttt{palettac@maths.tcd.ie}}, Anton Pribytok${}^*$\footnote{\texttt{apribytok@maths.tcd.ie}}, Ana L. Retore${}^*$\footnote{\texttt{retorea@maths.tcd.ie}} and Alessandro Torrielli${}^\dagger$\footnote{\texttt{a.torrielli@surrey.ac.uk}}}

\vspace{2em}

\vspace{1em}
\begingroup\itshape
${}^*$ School of Mathematics $\&$ Hamilton Mathematics Institute,
Trinity College,
Dublin, Ireland\\
${}^\dagger$ Department of Mathematics, University of Surrey, Guildford, GU2 7XH, UK
\par\endgroup

\end{center}

\vspace{2em}

\begin{abstract}\noindent 
In this paper we first demonstrate explicitly that the new models of integrable nearest-neighbour Hamiltonians recently introduced in PRL 125 (2020) 031604 \cite{MCAAP} satisfy the so-called {\it free fermion} condition. This both implies that all these models are amenable to reformulations as free fermion theories, and establishes the universality of this condition. We explicitly recast the transfer matrix in free fermion form for arbitrary number of sites in the 6-vertex sector, and on two sites in the 8-vertex sector, using a Bogoliubov transformation. We then put this observation to use in lower-dimensional instances of $AdS/CFT$ integrable $R$-matrices, specifically pure Ramond-Ramond massless and massive $AdS_3$, mixed-flux relativistic $AdS_3$ and massless $AdS_2$. We also attack the class of models akin to $AdS_5$ with our free fermion machinery. In all cases we use the free fermion realisation to greatly simplify and reinterpret a wealth of known results, and to provide a very suggestive reformulation of the spectral problem in all these situations.

\end{abstract}

\newpage

\overfullrule=0pt
\parskip=2pt
\parindent=12pt
\headheight=0.0in \headsep=0.0in \topmargin=0.0in \oddsidemargin=0in

\vspace{-3cm}
\thispagestyle{empty}
\vspace{-1cm}

{\hypersetup{linkcolor=black}
\tableofcontents
}

\setcounter{footnote}{0}

\section{Introduction}

\subsection{Integrability in $AdS_3$ and $AdS_2$ backgrounds}

Integrability of the $AdS_3 \times S^3 \times S^3 \times S^1$ and $AdS_3 \times S^3 \times T^4$ string-theory backgrounds  \cite{Bogdan,Sundin:2012gc} (see also \cite{rev3,Borsato:2016hud}), has proceeded taking the moves from the infinite-length treatment of the $AdS_5/CFT_4$ spin-chain. A substantial body of work is now available \cite{OhlssonSax:2011ms,seealso3,Borsato:2012ss,Borsato:2013qpa,Borsato:2013hoa,Rughoonauth:2012qd,PerLinus,CompleteT4,Borsato:2015mma,Beccaria:2012kb,Sundin:2013ypa,Bianchi:2013nra,Bianchi:2013nra1,Bianchi:2013nra2}. The novelty of massless modes imposes a new framework \cite{Sax:2012jv,Borsato:2016xns,Sax:2014mea,Baggio:2017kza} eventually echoing the connection between massless $S$-matrices and $CFT$s \cite{Zamol2,Fendley:1993jh,DiegoBogdanAle} and going beyond the reach of perturbation theory \cite{Lloyd:2013wza,Abbott:2012dd,Abbott:2014rca,MI,Abbott:2020jaa} (see also \cite{Eberhardt:2017fsi,Gaber1,Gaber2,Gaber3,MCAAP,Gaber4,Gaber5,
GaberdielUltimo,Prin,Prin1,Abbott:2015mla,Per9,Hoare:2018jim,AntonioMartin,Regelskis:2015xxa,JuanMiguelAle}).

The massless sector displays in a very clear fashion the nature of the string integrable model as a quantum-group deformation of two-dimensional Poincar\'e supersymmetry. The idea goes back to \cite{CesarRafa,Charles} in the five-dimensional case, revisited by the more recent \cite{Riccardo}. In the massless $AdS_3$ case one can go further \cite{JoakimAle,BorStromTorri,Andrea} to the point where one can show \cite{gamma1,gamma2} that there exists a change of variables which puts the massless non-relativistic $S$-matrix and its dressing factor in difference-form, in fact the exact same difference-form as the (non-trivial) BMN limit \cite{DiegoBogdanAle}. The massless TBA of \cite{DiegoBogdanAle} is then almost straightforwardly adapted to the full (massless) non-relativistic case \cite{gamma2}. 

The $\AdS_3$ background allows for a mixed-flux extension \cite{Cagnazzo:2012se,s1,s2,Babichenko:2014yaa,seealso12} which modifies the traditional magnon dispersion relation \cite{ArkadyBenStepanchuk,Lloyd:2014bsa}:
\begin{equation}
\label{eq:disp-rel}
E = \sqrt{\Big(m + \frac{k}{2 \pi} p\Big)^2 + 4 h^2 \sin^2\frac{p}{2}},
\end{equation}
where $k$ (which we will restrict to the natural numbers) stands for the WZW level, $m$ for the mass and $h$ for the coupling constant - see \cite{OhlssonSax:2018hgc} for a full study of the moduli space of this model. A particular relativistic limit has been studied in \cite{gamma2}, where another instance of nontrivial scattering for right-right and left-left moving modes was found, leading to a family of $CFT$s with exact TBA description. An important wealth of work on this and similar deformations has now been done \cite{Baggio}.

Finally a rather extreme example of $AdS/CFT$ integrability is the $AdS_2 \times S^2 \times T^6$ background \cite{ads2}, dual either to a superconformal quantum mechanics or to a chiral $CFT$ \cite{dual,gen}. The coset action is of the
Metsaev-Tseytlin form \cite{Metsaev:1998it,sc} for the quotient
$$
\frac{PSU(1,1|2)}{ SO(1,1) \times SO(2)}.
$$
The theory admits a $\mathbb{Z}_4$ automorphism \cite{Bena:2003wd} which is key to its classical integrability \cite{Sorokin:2011rr,Cagnazzo:2011at}. In \cite{Hoare:2014kma} an $S$-matrix was derived utilising a centrally-extended $\mathfrak{psu}(1|1)^2$ algebra built upon the BMN ground state
\cite{Berenstein:2002jq,amsw}. A similar series of steps were taken following the higher-dimensional recipes and comparing with the available perturbative results \cite{amsw}. The $AdS_2$ massive modes sit in a {\it long} representation \cite{Arutyunov:2009pw}, and only the massless modes are in {\it short} ones. The Yangian symmetry was studied in \cite{Hoare:2014kma,Hoare:2014kmaa}. It is known that comparison with perturbation theory is problematic for massless modes \cite{Per6, Per10, MI}, which is a signal that massless $S$-matrices are fundamentally distinct from massive ones \cite{Borsato:2016xns} - they rather describe certain massless renormalisation group flows between conformal field theories \cite{Zamol2}. 

The BMN limit for right-right and left-left movers is once again non-trivial \cite{Andrea2} and reminiscent of, although rather distinct from, ${\cal{N}}=1$ supersymmetric models \cite{Fendley:1990cy}. The $S$-matrix has an XYZ/8-vertex structure \cite{Baxter:1972hz, Baxter:1972hz1,Schoutens,MC}. The absence of a reference state prevents the familiar algebraic Bethe ansatz approach \cite{Levkovich-Maslyuk:2016kfv}, see also \cite{Nepo, Nepo1, Nepo2, Nepo3, Nepo4, Nepo5, Nepo6}. The strategy of \cite{Andrea2} relies on the {\it free fermion} condition \cite{MC,Ahn} which will be crucial to this paper as well, and on {\it inversion relations} \cite{Zamolodchikov:1991vh}. The transfer matrix was explicitly calculated up to 5 particles in \cite{Ale}, where a conjecture for the massless Bethe ansatz was given and for part of the massive Bethe equations. The free fermion condition holds in fact for the massive $AdS_2$ $S$-matrix as well, which we will show can be understood in the light of the results of this paper. This will provide a framework where to revisit the observations of \cite{amsw,Per6,Per10,Faddeev:1995nf,Faddeev:1995nf1} as well.

\subsection{This paper}

In this paper we will embed the above-mentioned integrable structures within the larger context of the classification performed in \cite{MCAAP}  - built upon the work in \cite{MAAP} - where a general study and a classification was made of integrable $R$-matrices and nearest-neighbour interaction spin-chain Hamiltonians of the 8- (or less) vertex type. Such classification includes the cases described in the previous subsection (with the exception of the massless $AdS_2$ situation which we will discuss in a separate subsection \ref{section53}). Here, we will show that the remarkable property which was observed for these $AdS$ models, namely the free fermion condition, holds in fact for the two new models introduced in \cite{MCAAP}. We will be giving plenty of details of this extraordinary unifying feature in the following sections, especially at the beginning of section \ref{analternativetotheABA}, where we will also provide a context for the associated literature. At this stage we remark that this confirms the universality of this condition and its far-reaching physical significance. It also furnishes an explanation of the diverse observations related to it which have been encountered so far within the lower-dimensional $AdS/CFT$ examples. \\
In section \ref{sec:freefermion}, we will give the mathematical proof of the fact that the two new models {({} {\it class B})} introduced in \cite{MCAAP} satisfy the free fermion condition, both by explicit verification \ref{explicitv}, and by use of the Sutherland equation \ref{suth}. The remaining models classified in \cite{MCAAP} satisfy instead a different ({\it Baxter}) condition - with both the free-fermion and the Baxter condition stemming from a universal functional relation which we prove in section \ref{sec:freefermion}. The free-fermion condition for class B models will allow us to use a diagonalisation procedure inspired by the coordinate Bethe ansatz, which sets the transfer matrix in manifest free fermion form for an arbitrary number of sites and inhomogeneities for the 6-vertex type $R$-matrices \ref{analternativetotheABA1}. The final results are remarkably compact and suggestive, and we consider them as a powerful alternative to the Algebraic Bethe Ansatz. In \ref{AdS3RR} we will then apply  this to the case of the pure Ramond-Ramond $AdS_3$ massless $R$-matrix and use the free fermion condition to revisit and drastically simplify the associated algebraic structure. An important point will be that the simplicity of this particular model will allow us to provide a very explicit display of our general diagonalisation formulas and a more detailed analysis of the specific associated algebraic structures. We will then follow a similar approach for the mixed-flux $AdS_3$ relativistic case in \ref{AdS3mixedf} and for the massive pure Ramond-Ramond $AdS_3$ case \ref{massiveAdS3}. 

In \ref{8vmodel}, we will then analyse the 8-vertex type models, which include the $AdS_2$ scattering problem (analysed in \ref{section53}). We will recast the two-site transfer matrix explicitly in a free fermion form, and in the process discover that there exists a state playing a role similar to that of the pseudo vacuum, which such models do not possess. This is because the particle-hole transformation we perform allows us to populate the spectrum in the usual fashion by adding excitations onto this state, which we therefore dub the {\it pseudo pseudo vacuum}. The construction is then paraphrased for massless $AdS_2$ excitations and the pseudo pseudo vacuum is shown. In \ref{sec:16}, we then apply the formalism to the case of $16 \times 16$ $R$-matrices and Hamiltonians exhibit $\mathfrak{su}(2)\oplus \mathfrak{su}(2)$ symmetry and manage to recast the latter in a form which is as close as possible to free fermions (achieving free fermions in special cases). This enters the territory of the Hubbard model and of the $AdS_5$ integrable scattering theory.  

In the appendices: in \ref{app:BA} we provide the proofs of the formulas we report in section \ref{analternativetotheABA} using ideas from the coordinate Bethe ansatz; in \ref{appB} we show how to derive from the general theory of section \ref{analternativetotheABA} the specific formulas for $AdS$; in \ref{appC} we obtain a bare recursive formula for the pure Ramond-Ramond massless $AdS_3$ transfer matrix on an arbitrary number of sites with inhomogeneities; in \ref{examples} we restrict to the homogeneous case and give explicit examples of the general formalism, namely the effective spin-chain Hamiltonian for pure Ramond-Ramond massless and mixed flux relativistic $AdS_3$, for open as well as closed chains, showing how the formulas simplify for the free fermion realisation to rather minimalistic expressions. 

Besides the general proof of the validity of this condition and its potential physical implications for these models, we hope that this paper will show the power of the free fermion realisation in organising the exact results for these systems in an astonishingly transparent and suggestive form. 

\section{Free fermion condition for general 8-vertex models}\label{sec:freefermion}

In \cite{MCAAP,MCAAP2}, building upon the earlier \cite{MAAP}, a classification was made of all the possible regular ({\it i.e.} nearest-neighbour) integrable spin-chains whose $R$-matrix admits the general 8-vertex form 
\begin{eqnarray}
R = \begin{pmatrix}r_1&0&0&r_8\\0&r_2&r_6&0\\0&r_5&r_3&0\\r_7&0&0&r_4 \end{pmatrix},
\label{Rmatrix}
\end{eqnarray}
where $r_i=r_i(u,v)$. It is the purpose of this section to verify that all integrable models of this type fall into two classes that are characterised by a general property of the $R$-matrix
\begin{equation}
\frac{(r_1r_4+r_2r_3-r_5r_6-r_7r_8)^2}{r_1r_2r_3r_4}=\text{const}.
\label{general(non)freefermion}
\end{equation}
It is then natural to consider two classes of models
\begin{itemize}
\item[ A] \textbf{Baxter condition}
\begin{equation}
\label{classA}
\text{const}\neq 0
\end{equation}
\item[ B] \textbf{The free fermion condition}
\begin{equation}
\text{const}=0
\end{equation}
which implies
\begin{eqnarray}
\label{free}
r_1 r_4 + r_2 r_3 = r_5 r_6 + r_7 r_8.
\end{eqnarray}
\end{itemize}
This classification was already known for models of difference form and we find that \eqref{general(non)freefermion} is the generalisation to models of non-difference form, which, in particular, include the AdS/CFT integrable models.

Class A corresponds to the models 6-vertex A and 8-vertex A from \cite{MCAAP}, while class B corresponds to the models 6-vertex and 8-vertex B from \cite{MCAAP}, as we check below. All holographic integrable models fall into class B.  Class A basically contains the usual XXZ and XYZ spin-chains. At the level of the Hamiltonian, the difference between the two models is the presence of the $S_z\otimes S_z$ interaction term in the Hamiltonian. Class A models contain this interaction, while Class B models do not. Interesting properties can be found in \cite{MartinsAndKhachatryan}.

In the next section we shall comment on the historical origin of free fermion condition and its use, and afterwards we shall use it to simplify specific models. In this section we shall prove that indeed the class B  solutions singled out in \cite{MCAAP} (modulo what we could call {\it gauge} transformations\footnote{See section \textit{Identifications} in \cite{MCAAP} for details.}) exactly satisfy the free fermion condition. We will then prove the above classification directly using the Sutherland equation, which allows for possible generalisations of the free fermion conditions to other types of models.

\subsection{Class A}

Up to basic identifications, models from class A correspond to the usual 6- and 8-vertex model. It is not hard to see that the condition \eqref{classA} is compatible with these identifications. Hence it remains to show that these models satisfy condition A. We will spell this out for the 8-vertex model and leave the 6-vertex model for the reader. The 8-vertex $R$-matrix is given in terms of Jacobi elliptic functions as
\begin{align}
R^{8v}(z) = \begin{pmatrix}
 \text{sn}(\gamma +z) & 0 & 0 & k\, \text{sn}(\gamma ) \text{sn}(z) \text{sn}(\gamma +z) \\
 0 & \text{sn}(z) & \text{sn}(\gamma ) & 0 \\
 0 & \text{sn}(\gamma ) & \text{sn}(z) & 0 \\
 k\, \text{sn}(\gamma ) \text{sn}(z) \text{sn}(\gamma +z) & 0 & 0 &  \text{sn}(\gamma +z) 
\end{pmatrix},
\end{align}
where $ \gamma $ and $ k $ are arbitrary constants. Thus, we find
\begin{align}
\frac{(r_1r_4+r_2r_3-r_5r_6-r_7r_8)^2}{r_1r_2r_3r_4} = 4 \text{cn}^2(\gamma ,k^2) \text{dn}^2(\gamma,k^2).
\end{align}
and indeed the Baxter condition holds.

\subsection{Class B} 
\label{explicitv}

The 6-vertex B $R$-matrix \cite{MCAAP,MCAAP2} is
\begin{align}
R = H_4(x,y)
\begin{pmatrix}
h_5(x) & 0 & 0 & 0 \\
0 & 1 & 0 & 0 \\
0 & 0 & h_5(x)h_5(y) & 0 \\
0 & 0 & 0 & -h_5(y) 
\end{pmatrix}
+
\begin{pmatrix}
1 & 0 & 0 & 0 \\
0 & 0 & 1 & 0 \\
0 & 1 & h_5(y) - h_5(x) & 0 \\
0 & 0 & 0 & 1 \\
\end{pmatrix},
\end{align}
where $ H_4(x,y)=H_4(x)-H_4(y) $; and $ H_4(x) $ and $ h_5(x) $ are arbitrary functions of $ x $.

One can then compute each term in \eqref{free} 
%
%
and obtain
\begin{equation}
r_1\,r_4+r_2\,r_3=1=r_5\,r_6+r_7\,r_8
\end{equation}
%
what proves that the free fermion condition \eqref{free} is satisfied for 6-vertex B.

For 8-vertex B the entries of \eqref{Rmatrix} are 
\begin{align}
r_1(u,v)&=\frac{1}{\sqrt{\sin\gamma(u)\sin\gamma(v)}}\left(\sin\gamma_+\frac{\text{cn}}{\text{dn}}-\cos\gamma_+ \text{sn}\right),\label{r1-8vB}\\ r_2(u,v)&=\frac{\pm 1}{\sqrt{\sin\gamma(u)\sin\gamma(v)}}\left(\cos\gamma_-\text{sn}+\sin\gamma_- \frac{\text{cn}}{\text{dn}}\right),\label{r2-8vB}\\
r_3(u,v)&=\frac{\pm 1}{\sqrt{\sin\gamma(u)\sin\gamma(v)}}\left(\cos\gamma_- \text{sn}-\sin\gamma_-\frac{\text{cn}}{\text{dn}}\right),\label{r3-8vB}\\
r_4(u,v)&=\frac{1}{\sqrt{\sin\gamma(u)\sin\gamma(v)}}\left(\sin\gamma_+\frac{\text{cn}}{\text{dn}}+\cos\gamma_+ \text{sn}\right),\label{r4-8vB}\\
r_5(u,v)&=r_6(u,v)=1,\quad r_7(u,v)=r_8(u,v)=k \frac{\text{sn}\,\text{cn}}{\text{dn}},\label{r5678-8vB}
\end{align}
\noindent
where $ \text{sn}=\text{sn}(u-v,k^2) $, $ \text{cn}=\text{cn}(u-v,k^2) $, $ \text{dn}=\text{dn}(u-v,k^2) $,  $\gamma_{\pm}\equiv\frac{\gamma(u)-\gamma(v)}{2}$, $k$ an arbitrary constant and $\gamma(u)$ an arbitrary function.
 
Let us start by the rhs of equation \eqref{free}.  
\begin{equation}
r_5r_6+r_7r_8=1+k^2 \frac{\text{sn}^2\,\text{cn}^2}{\text{dn}^2}.
\label{rhsfree8vertexB}
\end{equation}
Now for the left hand side of equation \eqref{free}, let us consider first $ r_1\,r_4 $
\begin{align}
r_1r_4& =\frac{1}{\sin\gamma(u)\sin\gamma(v)}\left(\sin^2\gamma_+\frac{\text{cn}^2}{\text{dn}^2}+\sin\gamma_+\cos\gamma_+\frac{\text{cn sn}}{\text{dn}}-\sin\gamma_+\cos\gamma_+\frac{\text{cn sn}}{\text{dn}}-\cos^2\gamma_+\text{sn}^2\right)\nonumber\\
&=\frac{\sin^2\gamma_+\frac{\text{cn}^2}{\text{dn}^2}-\cos^2\gamma_+\text{sn}^2}{\sin\gamma(u)\sin\gamma(v)}
\end{align}
where the two crossing terms cancelled.
Now let us compute $ r_2\,r_3 $
\begin{align}
r_2r_3& =\frac{1}{\sin\gamma(u)\sin\gamma(v)}\left(\cos^2\gamma_-\text{sn}^2-\sin\gamma_-\cos\gamma_-\frac{\text{cn sn}}{\text{dn}}+\sin\gamma_-\cos\gamma_-\frac{\text{cn sn}}{\text{dn}}-\sin^2\gamma_-\frac{\text{cn}^2}{\text{dn}^2}\right)\nonumber\\
&=\frac{\cos^2\gamma_-\text{sn}^2-\sin^2\gamma_-\frac{\text{cn}^2}{\text{dn}^2}}{\sin\gamma(u)\sin\gamma(v)}
\end{align}
which brings us to
\begin{align}
r_1r_4+r_2r_3&=\frac{\sin^2\gamma_+\frac{\text{cn}^2}{\text{dn}^2}-\cos^2\gamma_+\text{sn}^2}{\sin\gamma(u)\sin\gamma(v)}+\frac{\cos^2\gamma_-\text{sn}^2-\sin^2\gamma_-\frac{\text{cn}^2}{\text{dn}^2}}{\sin\gamma(u)\sin\gamma(v)}\nonumber\\
&=\frac{1}{\sin\gamma(u)\sin\gamma(v)}\left[\left(\sin^2\gamma_+-\sin^2\gamma_-\right)\frac{\text{cn}^2}{\text{dn}^2}-\left(\cos^2\gamma_+-\cos^2\gamma_-\right)\text{sn}^2\right]\nonumber\\
&=\frac{\text{cn}^2}{\text{dn}^2}+\text{sn}^2.
\label{lhsfree8vertexB}
\end{align}
%
%
%
Putting \eqref{rhsfree8vertexB} and \eqref{lhsfree8vertexB} together we have
\begin{align}
& \frac{\text{cn}^2}{\text{dn}^2}+\text{sn}^2=1+k^2 \frac{\text{sn}^2\,\text{cn}^2}{\text{dn}^2},\label{8vBfreeline1}\\
& 1-\text{sn}^2=\frac{\text{cn}^2}{\text{dn}^2}-k^2 \frac{\text{sn}^2\,\text{cn}^2}{\text{dn}^2},\label{8vBfreeline2}\\
& 1-\text{sn}^2=\frac{\text{cn}^2}{\text{dn}^2}\left(1-k^2\text{sn}^2\right),\label{8vBfreeline3}\\
& 1 = \text{cn}^2+\text{sn}^2,\label{8vBfreeline4}\\
& 1 = 1\label{8vBfreeline5}
\end{align}
what proves that 8-vertex B $R$-matrix satisfies the free fermion condition. Notice that from \eqref{8vBfreeline3} to \eqref{8vBfreeline4} we used the identity $ 1-k^2\text{sn}^2=\text{dn}^2 $ and from \eqref{8vBfreeline4} to \eqref{8vBfreeline5} we used $ 1 = \text{cn}^2+\text{sn}^2 $.

\subsection{Sutherland equation}
\label{suth}
One can in fact prove that any $R$-matrix of the form \eqref{Rmatrix} satisfies the generalised condition \eqref{general(non)freefermion} by using the Sutherland equations
\begin{align}
& \left[R_{13}(u,v)R_{23}(u,v),H_{12}(u)\right]=\dot{R}_{13}(u,v)R_{23}(u,v)-R_{13}(u,v)\dot{R}_{23}(u,v),\nonumber\\
&\left[R_{13}(u,v)R_{12}(u,v),H_{23}(v)\right]  =  R_{13}(u,v)R^\prime_{12}(u,v) - R^\prime_{13}(u,v)R_{12}(u,v),
\label{Sutherland}
\end{align}
where the dot and prime denote derivative with respect to $u$ and $v$, respectively.

Let us assume a Hamiltonian density of the form
\begin{equation}
\mathcal{H}=\begin{pmatrix}
h_1&0&0&h_8\\0&h_5&h_3&0\\0&h_2&h_6&0\\h_7&0&0&h_4
\label{Hamil8vertex}
\end{pmatrix}
\end{equation}
and an $R$-matrix of the form \eqref{Rmatrix}, with $ r_i $ being any function of $ u $ and $ v $ and $ h_i $ being any function of $ u $. 
In order to prove the relation \eqref{general(non)freefermion}, we start by doing the following procedure
\begin{enumerate}
\item Substitute \eqref{Rmatrix} and \eqref{Hamil8vertex} into the Sutherland equations \eqref{Sutherland}.
\item Solve for the derivatives $\dot{r_i}$ and $r_i'$ in terms of $ h_i $ and $ r_i $ (without actually solving the differential equations).
\item Then solve for some of the $ h_i $'s in terms of $r_i(u,v) $.
\end{enumerate}
Remarkably, by doing this, one obtains the following set of conditions on the coefficients of the $R$-matrix
\begin{align}
& \frac{r_1r_4+r_2r_3-r_5r_6-r_7r_8}{r_2r_4}=f(u),\label{FFBaxterpart1a}\\
& \frac{r_1r_4+r_2r_3-r_5r_6-r_7r_8}{r_1r_3}=g(u)\label{FFBaxterpart2a}\\
& \frac{r_1r_4+r_2r_3-r_5r_6-r_7r_8}{r_3r_4}=g(v),\label{FFBaxterpart1}\\
& \frac{r_1r_4+r_2r_3-r_5r_6-r_7r_8}{r_1r_2}=f(v)\label{FFBaxterpart2}
\end{align}
where $ f(u) $ and $ g(u) $ are functions of the matrix elements $ h_i(u) $ of the density Hamiltonian $ \mathcal{H} $ \eqref{Hamil8vertex}.

By multiplying \eqref{FFBaxterpart1a} and \eqref{FFBaxterpart2a} and then also \eqref{FFBaxterpart1} and \eqref{FFBaxterpart2}  we can see that 
\begin{align}
 \frac{(r_1r_4+r_2r_3-r_5r_6-r_7r_8)^2}{r_1r_2r_3r_4} = f(u)g(u) = f(v)g(v),
\end{align}
and hence proving that equation \eqref{general(non)freefermion} holds. Since the Sutherland equations \eqref{Sutherland} are obtained as a derivative of the Yang-Baxter equation, we proved that all integrable models of the form  \eqref{Rmatrix} belong to one of the two classes A and B.


Notice that this recipe can be applied to other models as well. We will work this out  in Section \ref{sec:16} for a special case of a $16\times16$ $R$-matrix and Hamiltonian with $\mathfrak{su}(2)\oplus \mathfrak{su}(2)$ symmetry. There we will use this method to demonstrate that the $\mathrm{AdS}_5$ $R$-matrix also satisfies similar free fermion conditions (cf. \cite{MitevEtAl}). Since all the $AdS_3$ and $AdS_2$ integrable models fall into generalised class B above, and $AdS_4$ relies on the $AdS_5$ $R$-matrix, we obtain the remarkable fact that all the $R$-matrices arising in the $AdS/CFT$ correspondence do satisfy associated free fermion condition\footnote{Generalised free fermion conditions for deformed and reduced symmetries in $ AdS_{5} $ and possible maps with the constructions based on \cite{Korepanov_1993a,Korepanov_1993b,Shiroishi:1995} are currently under investigation.}.

\section{An alternative to the Algebraic Bethe Ansatz}
\label{analternativetotheABA}
As we have demonstrated, all the integrable cases we shall discuss have a common feature: the associated $R$-matrices satisfy the so-called {\it free fermion} condition, whose role in the context of $AdS/CFT$ was highlighted in \cite{MitevEtAl}. Therefore, according to the standard lore, they are amenable to a description in terms of free fermions. Further analysis has been performed in \cite{Maillard:1996tc,Claude-M.:1997} also in the context of inversion relations, tetrahedral algebras and higher-dimensional vertex models. An extensive investigation of the free fermion realisation for a variety of models with different lattice geometries and symmetries can be found for instance in \cite{wheeler2011free}. Further work can be found in \cite{FeFree1,FeFree2,Mel}.

This idea and its realisation goes back to old literature \cite{Lieb,Lieb2,Ahn,Felderhof,Bazhanov,Baxter1} (more recent work has appeared in \cite{CrampeRafaVinet}). The papers \cite{Lieb,Lieb2} were able to rewrite the Hamiltonians and transfer matrices of the XY chain and of Ising-type chain and lattice models in a form which manifestly displays their free fermion nature. In order to achieve this, a particular transformation of the canonical spin-chain operators is performed. Inspired by those ideas, in this paper we will find the appropriate canonical transformations for the various situations we shall study. In doing so we will of course crucially rely on the fact which we have proven in the previous section, that all the $R$-matrices we are going to analyse satisfy the free fermion condition.

We will show explicitly in each case how to map the problem to one of free fermions, by displaying the transformation at the level of the transfer matrix, where everything can be made completely manifest.

We will mainly focus on the cases in which the spaces we deal with are two-dimensional, and are spanned by one boson $|\phi\rangle$ and one fermion $|\psi\rangle$. We introduce the following notation, suggestive of the treatment to ensue:
\begin{eqnarray}
|\phi\rangle \equiv |0\rangle, \qquad |\psi\rangle \equiv c^\dagger |0\rangle, \qquad c|0\rangle = 0,
\end{eqnarray}   
where we have introduced canonical fermionic creation and annihilation operators $c^\dagger$ and $c$, respectively, such that
\begin{eqnarray}
\{c,c^\dagger\}=1, \qquad \{c,c\} = \{c^\dagger,c^\dagger\}=0.
\end{eqnarray}
Later on, whenever we will need to work with multiple spaces, we shall introduce operators
 \begin{eqnarray}
\{c_i,c_j^\dagger\}=\delta_{ij}, \qquad \{c_i,c_j\} = \{c_i^\dagger,c_j^\dagger\}=0.
\label{canonicalmanysites}
\end{eqnarray} 
For example, $i,j$ may be ranging from $0$ to $2$ for the transfer matrices with one auxiliary $ \{ 0 \} $ and two physical spaces $ \{ 1,2 \} $. In general, they will be ranging from $0$ to $N$ (length of the spin chain). 

We notice that the graded tensor product of spaces is nicely encoded in these operators, and provides a more compact way of performing many of the Algebraic Bethe Ansatz manipulations. For instance, suppose we were to calculate the action of a matrix on the pseudovacuum such as 
\begin{eqnarray}
[E_{21} \otimes E_{21}] |\phi\rangle \otimes |\phi\rangle,
\end{eqnarray}  
where $E_{ab}$ are the matrices with all zeroes but $1$ in row $a$ column $b$, and states are numbered as $|1\rangle = |\phi\rangle$ and $|2\rangle =|\psi\rangle$ as usual, so that $E_{ab} |c\rangle = \delta_{bc} |a\rangle$. In the language of creation and annihilation operators we have
\begin{eqnarray}
[E_{21} \otimes E_{21}] |\phi\rangle \otimes |\phi\rangle = c_1^\dagger c_2^\dagger |0\rangle,
\end{eqnarray}
where $|0\rangle$ is used to denote now $|\phi\rangle \otimes |\phi\rangle$ by abuse of notation. The fermionic nature of the operators implies that 
\begin{eqnarray}
c_1^\dagger c_2^\dagger |0\rangle = - c_2^\dagger c_1^\dagger |0\rangle,
\end{eqnarray}
which is the equivalent of\footnote{We will sometimes write $\mathds{1}$ and sometimes simply $1$ to denote the identity operator. The context will always make it clear.} 
\begin{eqnarray}
[E_{21} \otimes \mathds{1}][\mathds{1} \otimes E_{21}] |\phi\rangle \otimes |\phi\rangle = - [\mathds{1} \otimes E_{21}][E_{21} \otimes \mathds{1}]|\phi\rangle \otimes |\phi\rangle.
\end{eqnarray}
In terms of a single space, the association is therefore
\begin{eqnarray}
c^\dagger = E_{21}, \quad c = E_{12}, \quad m \equiv c \,c^\dagger = E_{11}, \qquad n \equiv c^\dagger c = E_{22}, 
\end{eqnarray}
and clearly 
\begin{eqnarray}
m + n = \mathds{1} = E_{11} + E_{22}.
\end{eqnarray}
Let us now work out some explicit examples on how the free fermion condition can be exploited to diagonalize spin-chains. 

\subsection{Set-up}

Let us now demonstrate how we can use the free fermion condition to diagonalize the transfer matrix by considering 6-vertex model whose $R$-matrix satisfies the free fermion condition. 

Thus, we consider an $R$-matrix of the form \eqref{Rmatrix} with $r_7=r_8=0$. It can be written in terms of oscillators in the following form
\begin{align}\label{eq:Rosc}
R^{(osc)}_{ij}(u,v) = r_1 m_i m_j + r_2 n_i m_j + r_3m_in_j - r_4 n_i n_j - r_5 c_i c^\dag_j+r_6 c^\dag_i c_j,
\end{align}
where we suppressed the explicit dependence of $r_i$ on $(u,v)$ and $i$ and $j$ indicate the spaces in which the operators are acting. Suppose the $R$-matrix is regular, \textit{i.e.} $R(u,u) = P$, with $ P $ being the graded permutation operator, then the Hamiltonian density is given by the logarithmic derivative of the
$R$-matrix
\begin{align}
\mathcal{H}_{ij} = P \partial_u R^{(osc)}_{ij}(u,v) \Big|_{v=u}.
\end{align}
If we denote the corresponding derivative coefficients as $h_i = \partial_u r_i$, then we find in particular
\begin{align}\label{eq:Hosc}
\mathcal{H}^{(osc)}_{12} = &\ 
h_1  + (h_6-h_1) n_2 - (h_1+h_5)n_1 - (h_1+h_4-h_5-h_6) n_1n_2  +  h_3  c^\dag_2 c_1 +h_2 c^\dag_1c_2   .
\end{align}
On the level of the Hamiltonian, the free fermion condition \eqref{free} imposes that\footnote{The extra signs come from grading, since grading was not considered in \eqref{free}.} $h_1+h_4-h_5-h_6 =0$, which eliminates the $n_1n_2$ term. 

More generally, the conserved charges are generated by taking logarithmic derivatives of the transfer matrix
\begin{align}
T_N({{} \theta_0},\vec{\theta}) = {{}\mbox{str}_0 \big[ R_{01} (\theta_0 , \theta_1) \ldots R_{0N} (\theta_0 , \theta_N) \big].}
\end{align}
The parameters $\theta_i$ are local inhomogeneities, which for holographic models corresponds to momenta of world-sheet excitations. Throughout the paper we will always denote the supertrace over the auxiliary space by $\mbox{str}_0$, associated with the spectral parameter $\theta_0$. In case all the rapidities coincide $\theta_i = \theta$, then the first logarithmic derivative corresponds to the nearest-neighbour Hamiltonian \eqref{eq:Hosc}. For generic inhomogeneities, all the conserved charges have interaction range $N$.

\subsection{Solving homogeneous spin-chains with Free Fermions\label{omogenee}}

Let us now consider the Hamiltonian \eqref{eq:Hosc} for the homogeneous spin-chain of lenght $N$. To this Hamiltonian we then apply our non-local free fermion transformation to diagonalize it. We note that it is enough to consider the one-particle sector since this will induce a canonical map between the oscillators ${c},{c}^\dag$ and a new set of operators whose purpose will be to recast the Hamiltonian in a manifest free-fermion form. We shall denote these new operators, to be determined shortly in formula (\ref{canonicaltransf}), as $\eta,\eta^\dag$. These new operators will still satisfy canonical anticommutation relations. Restricted to this subsector, our Hamiltonian takes the simple form
\begin{align}
\mathbb{H}^{(1pt)} = \begin{pmatrix}
h_6+h_5 & h_2 & 0 & 0 &\hdots & 0 & h_3 \\
h_3 & h_6+h_5 & h_2 & 0 &\hdots & 0 & 0\\
0 & h_3 & h_6+h_5 & h_2 &\hdots & 0 & 0\\
0 & 0 & h_3 & h_6+h_5 & \hdots & 0 & 0\\ 
\vdots & \vdots & \vdots &\vdots & \ddots & \vdots & \vdots\\
0 & 0 & 0 & 0 & \hdots & h_6+h_5 & h_2\\
h_2 & 0 & 0 & 0 & \hdots & h_3 & h_6+h_5
\end{pmatrix}.
\end{align}
The eigenvalues and eigenvectors of $\mathbb{H}^{(1pt)}$ can now be easily computed. Let ${{} z}^N = 1$, then the eigenvectors {{} $\vec{\ell}$} and eigenvalues $\lambda$ are of the form
\begin{align}
&\lambda = h_6+h_5 + h_2 {{} z} + h_3{{} z}^{-1} ,
&& {{} \vec{\ell}} = (1,{{} z},{{} z}^2,\ldots {{} z}^{N-1}).
\end{align}
This means that there are exactly $N$ eigenvectors parameterised by the $N$th roots of unity and we can write the canonical transformation using $z=e^{\frac{2 \pi i k}{N}}$ for $k=1,...,N$
\begin{align}
&c_k  = \frac{1}{\sqrt{N}} {{} \sum_{n=1}^N} e^{2\pi i \frac{k n}{N} } \eta_n,
&&c_k^\dag  = \frac{1}{\sqrt{N}} {{} \sum_{n=1}^N} e^{-2\pi i \frac{k n}{N} } \eta_n^\dag.
\label{canonicaltransf}
\end{align}
This results in the Hamiltonian
\begin{align}
\mathbb{H} = h_1 N + \sum_{n=1}^N \Big[ (h_2 +h_3) \cos \frac{2\pi n }{N}+ i (h_2 - h_3) \sin \frac{2\pi n }{N}  -h_1+h_4\Big] \eta^\dag_n\eta_n,
\end{align}
which is now manifestly diagonal. Moreover, we can show that this canonical transformation also diagonalizes the full transfer matrix. Remarkably, only from $N=4$, we do need the free fermion condition for this. We find
\begin{align}
T_N = -\prod_{k=0}^{N-1}\bigg[ (r_1 - r_3 e^{\frac{2\pi i k}{N}}) M_{k+1} + (r_2 + r_4 e^{\frac{2\pi i k}{N}})  N_{k+1} \bigg],
\end{align}
where
\begin{align}
\label{asdefin}
&N_i=\eta^\dagger_i \eta_i,
&&M_i= \eta_i \eta^\dagger_i.
\end{align}
The transfer matrix takes a factorised form, which reminds of separation of variables. Because of this factorised form, we can exponentiate it and read off the conserved charges
\begin{align}
T_N =- \exp \Big[{{} i \psi_N} +  \sum_k {{} i \omega_k} N_k\Big],
\end{align}
where 
\begin{align}
&{{} i \psi_N} = \log (r_1^N - r_3^N),
&& {{} i \omega_k}= \log\bigg[\frac{r_2 + e^{\frac{2\pi i k}{N}}r_4}{r_1 - e^{\frac{2\pi i k}{N}}r_3}\bigg].
\end{align}
Since all the number operators commute, the exponent is well-defined.

\subsection{Inhomogenous spin-chains\label{analternativetotheABA1}}

Next we consider the case in which each spin-chain site has a corresponding inhomogeneity parameter $\theta_i$. In this case the one-magnon states are described by the inhomogeneous version of the Bethe Ansatz described in appendix \ref{app:BA}.

Similar to the homogeneous case, we construct the free fermion map by using the one-magnon eigenstates. We only need to invert the relation between the Bethe states and the basis vectors. The one-magnon states correspond to the $N$ solutions of the inhomogeneous Bethe equations
\begin{align}
\label{auxo}
1= \prod_{m=1}^N  S_m (v)  = \prod_{m=1}^N  \frac{r_1(\theta_m,v)}{r_2(\theta_m,v)}  .
\end{align}
Let us label the Bethe roots as $v_n$, then we obtain the following map
\begin{align}
&c^\dag_i =  {{} \sum_{n=1}^N} \frac{ f_i (v_n)\prod_{r=1}^{i-1} S_r (v_n)}{\sqrt{\sum_{r=1}^N |f_r(v_n)|^2}} \eta^\dag_n,
&&c_i =  {{} \sum_{n=1}^N} \frac{ f^*_i (v_n)\prod_{r=i}^{N} S_r (v_n)}{\sqrt{\sum_{r=1}^N |f_r(v_n)|^2}} \eta_n,
\label{inhocano}
\end{align}
where the expression for $f$ can be found in Appendix \ref{app:BA}.
We still denote the new operators with the symbols $\eta$ and $\eta^\dagger$, although the map is now different from the homogeneous case, therefore the explicit expression of the new oscillators  in terms of the original (and unchanged) $c$ and $c^\dagger$ will be different\footnote{{} Had we opted for explicitly indicating the dependence of the $\eta_n$ and $\eta^\dagger_n$ on the inhomogeneities $\theta_i$, we would then have that the $\eta$ operators defined by (\ref{canonicaltransf}) are obtained from the $\eta$ operators defined in (\ref{inhocano}) by setting all the inhomogeneities $\theta_i=0 \, \, \forall i=1,...,N$ in their argument.}.

It is then straightforward to check, at least for small lengths, that under this map all terms in the transfer matrix are again given by simple products of the number operators. In fact we again find the following factorised form of the transfer matrix
\begin{align}\label{eq:Tdiag}
T_N(\theta_0) = - \exp\Big[ {{} i \psi_N{{}(\theta_0)}} + \sum_{m=1}^N {{} i \omega_m}{{}(\theta_0)} N_m\Big],
\end{align}
with
\begin{align}
\label{gener}
&{{} i \psi_N{{}(\theta_0)}} = \log\bigg[\prod_{m=1}^N r_1({{}\theta_0}, \theta_m) - \prod_{m=1}^N r_3({{}\theta_0}, \theta_m) \bigg],
&&{{} i \omega_m}{{}(\theta_0)} = \log\bigg[\frac{r_3\left({{}\theta_0} ,v_m\right)}{r_1\left({{}\theta_0} ,v_m\right)}\bigg].
\end{align}
where we have here explicitly indicated the spectral parameter ${{}\theta_0}$ to distinguish it from the inhomogeneities $\theta_i$, $i=1,...,N$. We will sometimes suppress this explicit dependence for lightness of notation. Concluding, by using our free fermion transformation, we were able to derive this very compact and elegant form of the transfer matrix.

\section{Applications to $AdS_3/CFT_2$}

Let us now apply the formalism of free fermions to the various integrable models that arise in the $AdS/CFT$ correspondence. We will focus in this section on $AdS_3$ models. For the kind of models relevant to the $AdS_3/CFT_2$ correspondence, which are all of the 6-vertex type, we have constructed in {{} section \ref{analternativetotheABA1}} a completely general formula which includes the inhomogeneities for an arbitrary number of frame particles ({\it sites} of the level-one transfer matrix in the nesting procedure). The scope of this subsection is to display explicitly, for the particular $AdS_3$ models considered in the literature, how the general formalism unfolds {{} in more detail, where it is possible to work out more minutely many of the implicit expressions outlined in section \ref{analternativetotheABA1}}.  

\subsection{$AdS_3$ with pure Ramond-Ramond flux}
\label{AdS3RR}
\subsubsection{Transfer matrix analysis}

We start by writing the $R$-matrix for the massless sector of the pure Ramond-Ramond $AdS_3$ integrable system in the oscillator formalism we have equipped ourselves with. We will use the relativistic variable $\theta$ everywhere, although one may simply replace {{} $\theta$ with the difference-form variable which was introduced in} \cite{gamma1,gamma2} and all the formulas will hold true for the complete non-relativistic massless theory. We will also traditionally denote
\begin{eqnarray}
a(\theta) \equiv \mbox{sech} \frac{\theta}{2}, \qquad b(\theta) \equiv \tanh \frac{\theta}{2}, \qquad a_{ij} \equiv a(\theta_i - \theta_j), \qquad b_{ij} \equiv b(\theta_i - \theta_j).
\end{eqnarray}

One has
\begin{eqnarray}
R_{12}(\theta) = \Big[\cosh \frac{\theta}{2} \, (m_1 m_2 - n_1 n_2) - \sinh \frac{\theta}{2} \, (m_1 n_2 - n_1 m_2) - c_1 c^\dagger_2 + c^\dagger_1 c_2\Big]. 
\end{eqnarray}
We have chosen not to write the overall normalisation factor since it will not play a role for the moment - we will reinstate it whenever necessary at a later stage. With this normalisation, we can directly compare with the $6$-vertex $R$-matrix in \cite{MitevEtAl}:
\begin{eqnarray}
a_M = d_M = \cosh \frac{\theta}{2}, \qquad b_M = - c_M = i \sinh \frac{\theta}{2},
\end{eqnarray} 
where $a_M,b_M,c_M,d_M$ are the parameters used in \cite{MitevEtAl}. We can immediately see that the free fermion condition is satisfied:
\begin{eqnarray}
\label{beloM}
a_M d_M - b_M c_M = 1.
\end{eqnarray}
The same conclusion follows from (\ref{free}) in the 6-vertex case $r_7=r_8=0$, since the translation of the parameters w.r.t. \cite{MitevEtAl} is given by $a_M = r_4, b_M = i r_2, c_M = i r_3, d_M = r_1, r_6 = r_5 = 1$.
The $R$-matrix degenerates to the graded permutation operator for $\theta=0$.  

This $R$-matrix is actually describing a nested Bethe ansatz, where the pseudovacuum made of all $|\phi\rangle$ is truly a level-one pseudovacuum, as opposed to the true BMN vacuum made of all $|Z\rangle$ as described for instance in \cite{Sax:2012jv}. It does not make sense to calculate the Hamiltonian from it for $AdS$ purposes, while it does of course in the approach of \cite{MCAAP} which is using these $R$-matrices as generating functions themselves of integrable systems. For $AdS$ scopes one should compute the transfer matrix 
\begin{eqnarray}
T_N = \mbox{str}_0 R_{01} (\theta_0 - \theta_1) ... R_{0N} (\theta_0 - \theta_N)
\end{eqnarray}
with full non-equal inhomogeneities $\theta_1,..,\theta_N$, as done in \cite{DiegoBogdanAle}. Moreover, the inhomogeneities need to be interpreted as momenta and they satisfy the so-called {\it momentum-carrying} Bethe equations.

Let us now explicitly work out the free fermion map for this holographic model. For clarity, we will just focus on two physical spaces $N=2$, for which the transfer matrix reads
\begin{eqnarray}
\label{unruly}
T_2 &=&\frac{1 - b_{01} b_{02}}{a_{01}a_{02}} \, (m_1 m_2 - n_1 n_2) + \frac{b_{01} - b_{02}}{a_{01}a_{02}} \, (m_1 n_2 - n_1 m_2) + c_1^\dagger c_2 - c_1 c_2^\dagger\nonumber\\
&=&\frac{1}{a_{12}} \mathds{1} - e^{-\frac{\theta_{12}}{2}} c_1^\dagger c_1 - e^{\frac{\theta_{12}}{2}} c_2^\dagger c_2 +c_1^\dagger c_2 - c_1 c_2^\dagger, \qquad \theta_{ij} = \theta_i - \theta_j.  
\end{eqnarray}
{{} One arrives at} the following canonical {{}transformation, reminiscent of the one employed in} \cite{Lieb,Lieb2}:
\begin{eqnarray}
\label{tend}
c_1 = \cos \alpha \, \eta_1 - \sin \alpha \, \eta_2, \qquad  c_2 = \sin \alpha \, \eta_1 + \cos \alpha \, \eta_2, \qquad \cot 2\alpha = \sinh \frac{\theta_{12}}{2} \in \mathbbmss{R},
\end{eqnarray}
having restricted ourselves to real inhomogeneities corresponding to physical momenta of the frame particles\footnote{The inverse transformation is given by
\begin{eqnarray}
\eta_1 = \cos \alpha \, c_1 + \sin \alpha \, c_2, \qquad  \eta_2 = -\sin \alpha \, c_1 + \cos \alpha \, c_2.
\end{eqnarray}}. 
It is not difficult to see that the new creation and annihilation operators still satisfy
 \begin{eqnarray}
\{\eta_i,\eta_j^\dagger\}=\delta_{ij}, \qquad \{\eta_i,\eta_j\} = \{\eta_i^\dagger,\eta_j^\dagger\}=0, \qquad i,j=1,2,
\end{eqnarray}
which results in the transformation has the following effect on the transfer matrix:
\begin{eqnarray}
\label{led}
T_2 = \cosh \frac{\theta_{12}}{2} \Big[\mathds{1} - 2 \eta_1^\dagger \eta_1\Big].
\end{eqnarray}
We could have added a piece $+ 0 \, \eta^\dagger_2 \eta_2$ inside the square bracket just to prove the point that the transformation we have introduced completely diagonalises the transfer matrix on two physical spaces, and explicitly shows the free fermion property.

{{} The transformation (\ref{tend}) can be obtained in two independent ways. One is the brute-force verification which has led to the free-fermion form (\ref{led}). The other is by utilising the general formulas \eqref{auxo}-\eqref{inhocano} we have previously derived. We provide a summary of this second way in appendix \ref{appB}.}

Let us verify that we obtained the expected eigenvectors and eigenvalues as in \cite{DiegoBogdanAle}. The eigenvalues are now clearly extremely easy to compute, as it is just a matter of adding free fermion energies. This reveals an additive structure to the eigenvalues of the transfer matrix which was otherwise hidden in the old formalism. Moreover, the eigenvectors are naturally normalised if the pseudovacuum is, thanks to the canonical nature of the oscillators.

Since both $\eta_1$ and $\eta_2$ still annihilate the pseudovacuum, it is clear that $|0\rangle = |\phi \rangle \otimes |\phi\rangle$ is an eigenstate, with eigenvalue $\cosh \frac{\theta_{12}}{2}$. 

Then, we have
\begin{eqnarray}
\eta_1^\dagger |0\rangle = \sin \alpha \Big[ c_2^\dagger |0 \rangle + \cot \alpha \,  c_1^\dagger |0 \rangle\Big]=\sin \alpha \Big[ |\phi\rangle \otimes |\psi\rangle + \cot \alpha \, |\psi\rangle \otimes |\phi\rangle\Big],
\end{eqnarray}
with eigenvalue $-\cosh \frac{\theta_{12}}{2}$. 

Afterwards, we have  
\begin{eqnarray}
\eta_2^\dagger |0\rangle = \cos \alpha \Big[ c_2^\dagger |0 \rangle - \tan \alpha \,  c_1^\dagger |0 \rangle\Big]=\cos \alpha \Big[ |\phi\rangle \otimes |\psi\rangle - \tan \alpha \, |\psi\rangle \otimes |\phi\rangle\Big],
\end{eqnarray}
with eigenvalue $\cosh \frac{\theta_{12}}{2}$. 

Finally, we have $\eta_1^\dagger \eta_2^\dagger |0\rangle = |\psi \rangle \otimes |\psi\rangle$, with eigenvalue $-\cosh \frac{\theta_{12}}{2}$.

We make the choice of branch $\cot \alpha = - e^{\frac{-\theta_{12}}{2}}$, so that the above exactly reproduces the result of \cite{DiegoBogdanAle} once a series of simplifications are worked out.

It is instructive to compare explicitly with the Algebraic Bethe Ansatz. The creation operator of the exact transfer matrix eigenstates is $B({{} v};\theta_1,\theta_2)$, evaluated on the solutions of the auxiliary Bethe equations
\begin{eqnarray}
\label{aux}
\prod_{i=1}^N b({{} v} - \theta_i) = 1.
\end{eqnarray}
where $N=2$ in this particular case. As explained in appendix \ref{appB}, these Bethe equations are exactly obtained from (\ref{auxo}) specialised to the $R$-matrix we are dealing with in this case. The attribute of {\it auxiliary} used here stems from our earlier comment of being one level into the nesting procedure, where the inhomogeneous transfer matrix corresponds to the first level of the Algebraic Bethe Ansatz built upon the BMN vacuum \cite{DiegoBogdanAle}.

In the case of two physical spaces there are only two solutions: ${{} v}=\pm \infty$, corresponding to two creation operators which are easily translated in our new formalism:  
\begin{eqnarray}
\label{using}
B(+\infty;\theta_1,\theta_2) \propto (1 - 2 n_1) \, c_2^\dagger + e^{\frac{\theta_{12}}{2}} c_1^\dagger (1 - 2 n_2), \qquad 
B(-\infty;\theta_1,\theta_2) \propto c_2^\dagger - e^{-\frac{\theta_{12}}{2}} c_1^\dagger.
\end{eqnarray} 
{{}It is easy to see that this translates into
\begin{eqnarray}
B(+\infty;\theta_1,\theta_2) \propto \frac{1}{\cos \alpha} (1-2 N_1) \eta_2^\dagger, \qquad B(-\infty;\theta_1,\theta_2)  \propto \frac{1}{\sin \alpha} \eta_1^\dagger,
\end{eqnarray}
where $N_1 = \eta_1^\dagger \eta_1$ as defined in (\ref{asdefin}), hence}
\begin{eqnarray}
B(+\infty;\theta_1,\theta_2) |0\rangle \propto \frac{1}{\cos \alpha} \eta_2^\dagger |0\rangle, \qquad B(-\infty;\theta_1,\theta_2) |0\rangle \propto \frac{1}{\sin \alpha} \eta_1^\dagger |0\rangle,
\end{eqnarray}
establishing the equivalence of the Algebraic Bethe Ansatz to our formalism in generating eigenstates of the transfer matrix. We can also see how the $B$ operators do not straightforwardly create normalised eigenstates, but it is simple now to just use the $\eta_i^\dagger$ which do.

We can notice a certain similarity of our approach with the $B_{\mbox{good}}$ strategy of \cite{bgood}, in the spirit of trying to find a $B$-type operator which automatically implements the nesting procedure of the Algebraic Bethe Ansatz. In fact, in the case of pure Ramond-Ramond massless $AdS_3$ we hope that we might be able in the future to explicitly determine the operators $\eta_i^\dagger$ for any value of $N$ without the intermediate step of solving the auxiliary Bethe equations, exactly as we have been able to determine them for $N=2$. Solving the auxiliary Bethe equations is still a necessary stage in our general solution of section \ref{analternativetotheABA1}, but the $AdS_3$ case might be sufficiently simple to be able to directly conjecture the operators which create (normalised) eigenstates of the transfer matrix for generic $N$.

We can proceed to complete the map with the remaining operators of the Algebraic Bethe Ansatz. This allows establishing a correspondence between our free fermion realisation and the RTT relations - somewhat in the spirit of the Holstein-Primakoff oscillator realisations of Lie algebras. To be more precise, we have access from our formalism only to the RTT relations for the $B$ (and, shortly, $C$) operators evaluated on the solutions to the auxiliary Bethe equations (we could call them {\it on shell} RTT relations). Using (\ref{using}) it is already possible to see for instance that 
\begin{eqnarray}
B(+\infty;\theta_1,\theta_2)B(-\infty;\theta_1,\theta_2) = B(-\infty;\theta_1,\theta_2)B(+\infty;\theta_1,\theta_2),
\end{eqnarray} 
(and clearly the same with both $+\infty$ and both $-\infty$), which is one of the simplest RTT relations in \cite{DiegoBogdanAle} evaluated on shell. Similarly, we can prove
\begin{eqnarray}
(A-D) B(\pm\infty;\theta_1,\theta_2) = \frac{a(\mp\infty)}{b(\mp\infty)} B(\mp\infty;\theta_1,\theta_2) (A-D) - \frac{1}{b(\mp\infty)} B(\pm\infty;\theta_1,\theta_2) (A-D),
\end{eqnarray}
where we have used the fact that the transfer matrix $A-D$ happens to be independent of $\theta_0$ for two physical spaces. This is also another RTT relation from \cite{DiegoBogdanAle}.

If we take a look at the $C$ operators on shell, we discovered that they translate into\footnote{{} We are indebted with Juan Miguel Nieto and the anonymous referee for pointing out an error in $C(-\infty;\theta_1,\theta_2)$ in the first version of the manuscript. We have also verified that the relationship between $B^\dagger$ and $C$ for the massless pure-RR $AdS_3$ case derived in \cite{JuanMiguelAle} is exactly verified by the formulas we have obtained at ${{} v} \to \pm \infty$ in terms of free fermions.} {{}
\begin{eqnarray}
C(+\infty;\theta_1,\theta_2) \propto \frac{1}{\cos \alpha} (1-2 N_1) \eta_2, \qquad C(-\infty;\theta_1,\theta_2) \propto \frac{1}{\sin \alpha} \, \eta_1. 
\end{eqnarray}}
They do satisfy 
\begin{eqnarray}
C(+\infty;\theta_1,\theta_2)C(-\infty;\theta_1,\theta_2) = C(-\infty;\theta_1,\theta_2)C(+\infty;\theta_1,\theta_2),
\end{eqnarray} 
(and clearly the same with both $+\infty$ and both $-\infty$). It is unclear at this stage whether at higher $N$ one should expect the map between the operators of the Algebraic Bethe Ansatz and the free oscillators to be progressively more and more complicated. In this respect, the operators $A$ and $D$ separately are already rather complicated at this stage:
\begin{eqnarray}
&&A = e^{\theta_0 - \frac{\theta_1}{2} - \frac{\theta_2}{2}} N_1 \, N_2 - \frac{1}{2} e^{- \frac{\theta_1}{2} - \frac{\theta_2}{2}}(e^{\theta_0} + e^{\theta_1} + e^{\theta_2})N_1 - \frac{1}{2}e^{\theta_0 - \frac{\theta_1}{2} - \frac{\theta_2}{2}} N_2 + \frac{1}{a_{01} a_{02}},\nonumber\\
&&D=A-T_2. 
\end{eqnarray}
{{} Ideally,} we expect that at higher $N$ the relationship between $B$ and $\eta_i^\dagger$ (and between $C$ and $\eta_i$) will always be one-to-one upon explicitly substituting the solutions for the auxiliary roots, since the $B$ operators create eigenstates as much as $\eta^\dagger_i$ do. What might happen is
\begin{eqnarray}
B(\mbox{solution $i$ to auxiliary Bethe eqs};\theta_1,..,\theta_N) = f(N_1,...,N_N) \, \eta_i^\dagger,
\end{eqnarray}
for some function $f$ of all the number operators, while for the $C$ operators we ideally expect to have
\begin{eqnarray}
C(\mbox{solution $i$ to auxiliary Bethe eqs};\theta_1,..,\theta_N) = g(N_1,...,N_N) \, \eta_i,
\end{eqnarray}
and for the $A$ and $D$ operators 
\begin{eqnarray}
A(\theta_0;\theta_1,..,\theta_N) = F(N_1,...,N_N),\qquad {{}D(\theta_0;\theta_1,..,\theta_N) = G(N_1,...,N_N)},
\end{eqnarray}
all of the above being for suitable functions of the number operators (and of the appropriate sets of rapidities). These functions are polynomials in the $N_j$, $j=1,..,N$, and rational functions in the variables $e^\frac{\theta_i}{2}$, $i=0,...,N$. It is an interesting and challenging mathematical question to determine these functions recursively using the RTT relations.

Let us now make some remarks on the generalisation to higher values of $N$. 
Can we see in the explicit eigenvalues of the $AdS_3$ transfer matrix the additive pattern of eigenvalues which a free fermion description seems to suggest? 
The answer is of course positive. Strong in the knowledge of the general formula we have found in section \ref{analternativetotheABA1}, we now proceed to analyse how this applies to our particular case.

The generic eigenvalue $\lambda$ of the transfer matrix $T_N$ reads \cite{DiegoBogdanAle} 
\begin{eqnarray}
\lambda = {{} \Xi_N}(\theta_0;\theta_1,...,\theta_N) \prod_{m=1}^M \frac{1}{b({{} v}_m - \theta_0)},
\end{eqnarray}
where the ${{} v}_m$ are $M$ values chosen amongst the $N$ possible solutions to the auxiliary Bethe equations (\ref{aux})
\begin{eqnarray}
\label{auxa}
\prod_{i=1}^N b({{} v} - \theta_i) = 1,
\end{eqnarray}
{{}where we restrict to the strip ${{} v} \in [-i\frac{\pi}{2},i\frac{\pi}{2}]$ in order not to overcount. As discussed in \cite{Fendley:1993pi}, due to the properties of the tanh function the solutions to (\ref{aux}) localise in two families precisely corresponding to ${{} v_j} = z_j \pm i \frac{\pi}{2}$, with $z_j$ being real (sometimes called the {\it centres} of the solutions). In appendix B.2 of \cite{DiegoBogdanAle} it is shown how these two families of solutions for the auxiliary Bethe roots exactly provide a one-to-one correspondence with the $2^N$ transfer matrix eigenstates.}

The factor ${{} \Xi_N}$ is a complicated function of the rapidities and of the spectral parameter, but it is common to all eigenvalues. What changes is the number of excitations $M$, specifically the different choices of $M$ solutions out of the $N$ possible solutions of (\ref{aux}), corresponding to how many magnons have been turned on and with which magnon-rapidities ${{} v}_m$. From the nature of (\ref{aux}) it is clear that solutions with finite ${{} v}$ come in pairs, the second solution always corresponding to inverting all the factors of $b$, equivalently sending ${{} v} \to {{} v} + i \pi$. There is always then ${{} v} = \infty$ solution, and finally there is the solution ${{} v}=-\infty$ for even $N$. 

This is consistent with the expression for the transfer matrix (\ref{eq:Tdiag}) - modulo an inconsequential redefinition of $\psi_N \to \psi_N + \pi$:
\begin{eqnarray}
\label{mani}
T_N = e^{i \psi_N \mathds{1} + i \sum_{i=1}^N \omega_i \, \eta_i^\dagger \eta_i},
\end{eqnarray}
(where we recall that there is no need of Baker-Campbell-Hausdorff given the mutual commutativity of all the different $N_i \equiv \eta_i^\dagger \eta_i$). We know that the complication is buried in the explicit form of the $\psi_N$, $\omega_i$ and $\eta_i$, $\eta^\dagger_i$. These quantities will depend on $\theta_0$ and $\theta_1,..,\theta_N$, except the creation and annihilation operators which  only depend on $\theta_1,..,\theta_N$ (this is because the eigenstates cannot depend on $\theta_0$ by the mutual commutativity of the transfer matrix taken at different values of the spectral parameter). 

{{} An important comment regarding the notation which we have decided to adopt: for each value of $N$, there will be $N$ independent pairs of operators $\eta_i,\eta^\dagger_i$, $i=1,...,N$, their explicit form also changing with $N$. This means that we should rather write $\eta_i(N),\eta^\dagger_i(N)$. With respect to their version appearing in the previous section, these operators are thought to be the result of having explicitly solved the auxiliary Bethe equations. If we suppress the explicit $N$-dependence is merely to lighten up the formulas, however this distinction is often crucial and we shall remark it whenever it is particularly useful.}

Conforming to our expectations from the structure of the Algebraic Bethe Ansatz, we can argue that one of the $\omega_i$ will always be zero, corresponding to the ${{} v}=\infty$ auxiliary rapidity of the Algebraic Bethe Ansatz. We could have decided to always make this to be $\omega_N=0$ and truncate the sum to $N-1$. For even $N$, another one of the $\omega_i$ will be equal to $\pi$. The remaining ones will come in pairs of the type $(\omega_k,-\omega_k)$.

The $N=2$ case is clearly recovered very easily: given the fermionic nature of the operators, we have the standard Fermi property
\begin{eqnarray}
N_i^2 = N_i, \qquad N_i = \eta^\dagger_i \eta_i, \qquad i = 1,...,N,
\end{eqnarray}
from which we obtain, at $N=2$ and with the choice $\omega_2=0$, $\omega_1 = \pi$, that
\begin{eqnarray}
\label{esi}
T_2 = e^{i \psi_2} \, e^{i \omega_1 N_1} = e^{i \psi_2} \, \Bigg[1+ N_1 \sum_{n=1}^\infty \frac{(i \omega_1)^n}{n!} \Bigg] = e^{i \psi_2} \Big[1+ N_1 (e^{i \omega_1} - 1) \Big] = e^{i \psi_2} \Big[1 - 2 N_1\Big].
\end{eqnarray}
Setting $\psi_2 = -i \log \cosh \frac{\theta_{12}}{2}$ reproduces the desired result. In fact we can also write in a remarkably compact way
\begin{eqnarray}
T_2 = \Big[\cosh \frac{\theta_{12}}{2}\Big] \, e^{i \pi N_1} = \Big[\cosh \frac{\theta_{12}}{2}\Big] \, (-)^{N_1}.
\end{eqnarray}

As a further test of this idea, we can check the $N=3$ case. From \cite{DiegoBogdanAle}, once the appropriate common factor has been taken out, we have three eigenvalues corresponding to the following three different $\omega_i$ in our ansatz:
\begin{eqnarray}
\label{eneg}
\omega_1 = 0, \qquad \omega_2 = i \log \frac{y^2 - \mu^2}{y^2 + \mu^2}\equiv \omega, \qquad \omega_3 = - i \log \frac{y^2 - \mu^2}{y^2 + \mu^2}= -\omega, 
\end{eqnarray}
where
\begin{eqnarray}
\label{eneg1}
\mu = e^{\frac{\theta_0}{2}}, \qquad y = - e^{- i \frac{\pi}{4}} e^{\frac{\theta_1 + \theta_2 + \theta_3}{4}} (e^{\theta_{1}} + e^{\theta_{2}} + e^{\theta_{3}})^{-\frac{1}{4}}.
\end{eqnarray}
The pattern is exactly what we expected from the general arguments of section \ref{analternativetotheABA1}\ \footnote{{}In particular, we have seen in appendix \ref{appB} that the auxiliary Bethe equations (\ref{aux}), upon which the three values (\ref{eneg}) have been based in \cite{DiegoBogdanAle}, exactly coincide with (\ref{auxo}), therefore the three inequivalent solutions to such equations for $N=3$ are exactly the same. When plugging in these three auxiliary roots $v_i$ into the general expression - second formula in (\ref{gener}), which reduces for our situation to $i \omega_i = \log \tanh \frac{\theta_0 - v_i}{2}$ - one exactly recovers the three values (\ref{eneg})-(\ref{eneg1}) with a global shift by a constant which redefines the normalisation $\psi_N$.}. However, the true test is whether the $\omega_i$ are all real. Indeed, we can see that
\begin{eqnarray}
\frac{y^2 - \mu^2}{y^2 + \mu^2} = \frac{e^{-\theta _0}-i e^{-\frac{1}{2} \left(\theta _1+\theta _2+\theta _3\right)} \sqrt{e^{\theta _1}+e^{\theta _2}+e^{\theta _3}}}{e^{-\theta _0}+i e^{-\frac{1}{2} \left(\theta _1+\theta _2+\theta
		_3\right)} \sqrt{e^{\theta _1}+e^{\theta _2}+e^{\theta _3}}}
\end{eqnarray} 
which is a pure phase. The transfer matrix (\ref{mani}) is then manifestly unitary (when suitably normalised).

Finally, a rather non-trivial check is whether the additive pattern is reproduced. According to our ansatz, the $8$ eigenstates of the transfer matrix on two physical spaces should have eigenvalues respecting the following scheme:
\begin{eqnarray}
\{e^0,e^0,e^\omega,e^{-\omega},e^{\omega},e^{-\omega},e^0,e^{0}\},
\end{eqnarray} 
obtained by creating at most three free fermions and adding the energies (\ref{eneg}). This is exactly the pattern one observes in \cite{DiegoBogdanAle} once the common factor is taken out.

{{} If our conjecture (\ref{mani}) did turn out to be true in the end, then, by the general commutativity of the $N_i$, we could} write 
\begin{eqnarray}
\label{fron}
{{} T_N = e^{i \psi_N} \prod_{i=1}^N \Big[1 + (e^{i \omega_i}-1) N_i\Big]},
\end{eqnarray}
which would be a dramatic simplification of the general result \eqref{eq:Tdiag} for the case of the RR AdS${}_3$ model. {{} We can see that in the case $N=2$ the result (\ref{esi}) we have previously obtained is recovered from (\ref{fron}) since $\omega_1 = \pi$ and $\omega_2 = 0$.}

\subsubsection{The $R$-matrix and its symmetries}

Remarkably, in the new variables we also have
\begin{eqnarray}
\label{surprise}
R_{12}(\theta_1 - \theta_2) = T_2 = \cosh \frac{\theta_{12}}{2} \Big[\mathds{1} - 2 \eta_1^\dagger \eta_1\Big] = \Big[\cosh \frac{\theta_{12}}{2}\Big] \, e^{i \pi N_1}.
\end{eqnarray}
{{} This statement and the rest of this subsection have to be understood as now focusing, for the sole purposes of the $R$-matrix, exclusively on two sites (in this instance sites $1$ and $2$) and taking as the definition of the variables $\eta_1$ and $\eta_2$ the very transformation (\ref{tend}) associated to these two particular sites. Equivalently, they are what we would call $\eta_1(2)$ and $\eta_2(2)$. We also keep denoting here $N_i = \eta_i^\dagger \eta_i$ as number operators\footnote{{} The form of the $R$-matrix in terms of the new oscillators can also be adapted in principle to an arbitrarily-chosen isolated pair of sites, with the associated $\eta$ variables being in that case given by the transformation (\ref{tend}) with the inhomogeneities adapted to those two specific sites. We will make use of this fact in appendix \ref{appC}, where we will shall also introduce an appropriate notation to distinguish these oscillators from those directly diagonalising the $N$-site transfer matrix $T_N$.}.}
Considering once again that we are counting fermions, hence
\begin{eqnarray}
\label{virtute}
N_1^2 = N_1,
\end{eqnarray}
then it is easy to recover \cite{DiegoBogdanAle,JoakimAle,Andrea} that $\tilde{R} = \frac{R}{\cosh \frac{\theta_{12}}{2}}$ is both hermitian and unitary for real rapidities, since
\begin{eqnarray}
\tilde{R}^\dagger \tilde{R} = \tilde{R}^2 =  (\mathds{1} - 2 N_1)(\mathds{1} - 2 N_1) = \mathds{1}
\end{eqnarray}
by virtue of (\ref{virtute}). This is also straightforward to realise in the form
\begin{eqnarray}
\label{prob}
\tilde{R} = e^{i \pi N_1} = (-)^{N_1}, \qquad N_1 = \eta_1^\dagger \eta_1.
\end{eqnarray}

We find the expression (\ref{prob}) probably to be the simplest possible way we can write the massless $AdS_3$ $R$-matrix. It makes most of its properties completely evident, included the fact that Hermitian unitary matrices are diagonalisable and have eigenvalues $\pm 1$. Projectors onto the corresponding eigenspaces can be obtained by considering $P_\pm = \frac{1 \pm \tilde{R}}{2}$. In appendix \ref{appendB} we elaborate on how such a (perhaps deceptively) simple form of the $R$-matrix, where the complication is buried in the definition of the operators $\eta_i$ and $\eta^\dagger_i$, may be used to generate a recursion relation for the transfer matrix which by-passes the intermediate step of the auxiliary Bethe equations. Here we focus instead on how the symmetries of the $R$-matrix manifest itself in the free fermion language. 

It is easy to see that one has the following supersymmetry property:
\begin{eqnarray}
Q \equiv e^{\frac{\theta_1}{2}} c_1 +e^{\frac{\theta_2}{2}} c_2, \qquad Q^\dagger \equiv e^{\frac{\theta_1}{2}} c_1^\dagger +e^{\frac{\theta_2}{2}} c_2^\dagger, \qquad [Q,R]=0=[Q^\dagger,R]
\end{eqnarray}  
($Q^\dagger$ being the hermitian conjugate of $Q$ for real rapidities), which can be proven using the free fermion anti-commutation relations. In the traditional Hopf-algebra language we were briefly recapitulating at the beginning of section \ref{analternativetotheABA}, this is nothing but the {\it coproduct} of the supercharges:
\begin{eqnarray}
&&\Delta(q) = q_1 \otimes \mathds{1} + \mathds{1} \otimes q_2, \qquad q_i = \sqrt{p_i} \, E_{12}, \qquad p_i = {{} e^{\theta_i}},\nonumber\\
&&\Delta(q^\dagger) = q^\dagger_1 \otimes \mathds{1} + \mathds{1} \otimes q^\dagger_2, \qquad q^\dagger_i = \sqrt{p_i} \, E_{12},
\end{eqnarray}
where we have introduced the momenta $p_i$, $i=1,2$. The conditions $[Q,R] = 0 = [Q^\dagger,R]$ are the translation of the coproduct conditions
\begin{eqnarray}
\Delta^{op}(q) R = R \Delta(q), \qquad \Delta^{op}(q^\dagger) R = R \Delta(q^\dagger),
\end{eqnarray}
where the {\it opposite coproduct} $\Delta^{op}$ is obtained by permuting (in a graded fashion) the two factors in the tensor product of {{} $\Delta$}. One also uses the fact that $\Delta(q) = \Delta^{op}(q)$ and $\Delta(q^\dagger) = \Delta^{op}(q^\dagger)$ to reduce these relations to simple commutators. 

The operation of taking the opposite coproduct is simply translated in free fermion language by the rule: {\it ``swap the indices of particle $1$ and $2$"}. Because of the fermionic nature of the creation and annihilation operators, this rule automatically accounts for any fermionic sign, as the following example shows:
\begin{eqnarray}
\big(c_1 c^\dagger_2\big)^{op} = c_2 c^\dagger_1 = - c^\dagger_1 c_2,
\end{eqnarray}  
which perfectly reproduces the graded rule
\begin{eqnarray}
\Big(E_{12} \otimes E_{21}\Big)^{op} = - E_{21} \otimes E_{12}.
\end{eqnarray}


This supersymmetry respects the difference form of the $R$-matrix, since it reduces to the basic conditions
\begin{eqnarray}
[Q_0,R]=0=[Q^\dagger_0,R], \qquad Q_0 = e^{\frac{\theta}{2}} c_1 + c_2, \qquad Q^\dagger_0 = e^{\frac{\theta}{2}} c_1^\dagger + c_2^\dagger, \qquad \theta = \theta_1 - \theta_2,
\end{eqnarray}
via eliminating a scalar function of the rapidities (namely, factoring out $e^\frac{\theta_2}{2}$). In the language of the $\eta_i$ and $\eta^\dagger_i$ operators, it is immediate to see that
\begin{eqnarray}
Q \propto \eta_2, \qquad Q^\dagger \propto \eta_2^\dagger.
\end{eqnarray}
It is therefore straightforward to check the invariance under supersymmetry, since the $R$-matrix commutes with (in fact any function of) $\eta_2$ and $\eta_2^\dagger$ by virtue of the commutation relations of these operators. This provides a noticeable reinterpretation of the supersymmetry coproduct: it is given by those canonical combinations of creation and annihilation operators on which the $R$-matrix trivially does not depend.

The ${}^{op}$ transformation has a consequence on the variable $\alpha$ as well, since it exchanges the labels $1$ and $2$. The effect is encoded in the following map:
\begin{eqnarray}
\alpha \to \frac{\pi}{2} - \alpha \qquad \mbox{under} \, \, \, {}^{op}. 
\end{eqnarray}
This means that 
\begin{eqnarray}
\label{usin}
\eta_1^{op} = \eta_1, \qquad \eta_2^{op} = - \eta_2.
\end{eqnarray}
The relation $Q^{op} = Q$ is easily seen by considering that
\begin{eqnarray}
Q = e^{\frac{\theta_2}{2}} \Big[e^{\frac{\theta}{2}} c_1 + c_2\Big] =  e^{\frac{\theta_2}{2}} \big[-\tan \alpha \,  c_1 + c_2\big] = \frac{e^{\frac{\theta_2}{2}}}{\cos \alpha} \, \eta_2, 
\end{eqnarray}
therefore
\begin{eqnarray}
Q^{op} = -\frac{e^{\frac{\theta_1}{2}}}{\sin \alpha} \, \eta_2 = e^{\frac{\theta_1}{2}} \, \big[c_1 - \cot \alpha \, c_2\big] = e^{\frac{\theta_1}{2}} \, \big[c_1 + e^{-\frac{\theta}{2}} \, c_2\big] = Q. 
\end{eqnarray}
Using (\ref{usin}) it is immediate to see that $\tilde{R}^{op} = \tilde{R}$ \cite{DiegoBogdanAle,Andrea}, and in fact also $R^{op} = R$. Since we have $\tilde{R}^2 = \mathds{1}$, this implies braiding unitarity:
\begin{eqnarray}
\tilde{R} \, \tilde{R}^{op} = \tilde{R}_{12}(\theta) \, \tilde{R}_{21}(-\theta) = \mathds{1}.
\end{eqnarray} 

A longer calculation along the same lines reproduces the result of \cite{DiegoBogdanAle,Andrea}:
\begin{eqnarray}
\label{equa}
\Big[\frac{d}{d\theta} - \frac{1}{2} \mbox{sech} \frac{\theta}{2} 
\, (c_1 c_2^\dagger + c_1^\dagger c_2)\Big]\tilde{R} = 0, \qquad \tilde{R} 
= \mbox{sech} \frac{\theta}{2} \, R = 1 - 2 N_1,
\end{eqnarray}
which allows the rewriting (cf. \cite{DiegoBogdanAle})
\begin{eqnarray}
\tilde{R} = \Pi_s \, e^{- 2 \big[\arctan \tanh \frac{\theta}{4} \big]\, (c_1 c_2^\dagger + c_1^\dagger c_2)},
\end{eqnarray}
$\Pi_s$ being the graded permutation acting on states. In verifying these formulas one has to be careful with the fact that the operators $\eta_i$ and $\eta^\dagger_i$ depend themselves on the rapidities via their definition. Equation (\ref{equa}) does not represent a symmetry of the $R$-matrix but rather a constraint. The boost symmetry in the Hopf-algebra sense has been discussed in \cite{Riccardo,Riccardo2}. There is of course the standard relativistic boost symmetry of the $R$-matrix which reads
\begin{eqnarray}
\label{equa2}
[J,R]=0, \quad J = J_1 + J_2 = \frac{\partial}{\partial \theta_1} + \frac{\partial}{\partial \theta_2} \quad \longrightarrow \, \, \, \partial_{\theta_1 + \theta_2} R=0, \quad \rightarrow R = R(\theta_1-\theta_2) = R(\theta).
\end{eqnarray}
In the case of non-relativistic massless excitations it took a non-trivial series of steps \cite{gamma1,gamma2} to recast the $R$-matrix in such a way that, in the appropriate variables, such symmetry became completely manifest. Combinations of the conditions (\ref{equa}) and (\ref{equa2}) can be taken to produce derivative conditions separately w.r.t. $\theta_1$ and $\theta_2$ in a variety of ways \cite{Andrea,gamma1}. We speculate that these two conditions, together with the other symmetries displayed in this section, are in fact a basis for all the (differential) relations satisfied by this $R$-matrix.

There is a more involved supersymmetry, which however is the natural one from the string theory viewpoint. In fact, as discovered in \cite{gamma1,gamma2}, the remarkable property of the massless (right-right and left-left moving) $R$-matrix of pure Ramond-Ramond $AdS_3$ string theory is that it looks exactly the same both in the non-relativistic and in the relativistic (BMN) limit, provided one simply adjusts the rapidity variable. In particular, difference form is always there in the appropriate variables. We can recover this fact here: if we take the combinations
\begin{eqnarray}
Q_\pm =  {{} \sigma_{12,\pm}} \, c_1 + {{} \rho_{12,\pm}} \, c_2,  \qquad {{} \sigma_{12,\pm}} =  \sqrt{\sin \frac{\tilde{p}_1}{2}} \, e^{\pm i \frac{\tilde{p}_2}{4}}, \qquad {{} \rho_{12,\pm}} = \sqrt{\sin \frac{\tilde{p}_2}{2}} \, e^{\mp i \frac{\tilde{p}_2}{4}},
\end{eqnarray}  
where we notice the appearance of the familiar non-local braiding of the $AdS$ coproduct \cite{B1}, we can impose that they satisfy
\begin{eqnarray}
\label{hermi}
(Q_\pm)^{op} R = R \, Q_\pm.
\end{eqnarray}
This condition is consistent with difference form, since it simply amounts to the two separate conditions
\begin{eqnarray}
&&{{} \sigma_{21,\pm}} \, e^{\frac{\theta}{2}} - {{} \rho_{21,\pm}} = {{} \sigma_{12,\pm}} - {{} \rho_{12,\pm}} \, e^{\frac{\theta}{2}}, \qquad \theta = \theta_1 - \theta_2,\nonumber\\
&&{{} \sigma_{21,\pm}} + {{} \rho_{21,\pm}} \, e^{\frac{\theta}{2}} = {{} \sigma_{12,\pm}} \, e^{\frac{\theta}{2}} + {{} \rho_{12,\pm}}.
\end{eqnarray}
It is easy to manipulate both of these equations into one and the same form  by simple use of trigonometric identities:
\begin{eqnarray}
\label{solves}
\sqrt{\frac{\tan \frac{\tilde{p}_2}{4}}{\tan \frac{\tilde{p}_1}{4}}} = e^{-\frac{\theta}{2}}.
\end{eqnarray}
It is now immediate to see that requiring \cite{gamma1}
\begin{eqnarray}
\theta_i = \log \tan \frac{\tilde{p}_i}{4}
\end{eqnarray}
solves (\ref{solves}). We can then simplify $Q_\pm$ and obtain
\begin{eqnarray}
\label{front}
Q_\pm = g_{12}\, \Big( Q \pm i \big(e^{\frac{\theta_1}{2} + \theta_2} c_1 - e^{\frac{\theta_2}{2} + \theta_1} c_2\big)\Big),
\end{eqnarray} 
where $g_{12}$ is a symmetric function under the exchange $1 \leftrightarrow 2$.
If we revisit the condition of invariance of the $R$-matrix under this symmetry, we notice that we can drop the symmetric function in front of (\ref{front}), because it is invariant under ${}^{op}$ and drops out from the equation. We can then also forget about $Q$ which is also ${}^{op}$ invariant and we have already verified that it is a symmetry. We are left with the condition
\begin{eqnarray}
\label{lab}
\big(e^{\frac{\theta_1}{2} + \theta_2} c_1 - e^{\frac{\theta_2}{2} + \theta_1} c_2\big)^{op} R = R \big(e^{\frac{\theta_1}{2} + \theta_2} c_1 - e^{\frac{\theta_2}{2} + \theta_1} c_2\big).
\end{eqnarray}
By dividing both sides by $e^{\frac{\theta_1}{2} + \theta_2}$, we see that this condition is nothing else then
\begin{eqnarray}
\big(c_1 - e^{\frac{\theta}{2}} c_2\big) \, R = R \, \big(- c_1 + e^{\frac{\theta}{2}} c_2), \qquad \theta = \theta_1 - \theta_2,
\end{eqnarray}
or equivalently 
\begin{eqnarray}
\label{rather}
\{R, c_1 - e^{\frac{\theta}{2}} c_2\}=0 \qquad \longrightarrow \qquad \{R,\eta_1\}=0=\{R,\eta^\dagger_1\}, 
\end{eqnarray}
which completely manifests the difference-form. It is also immediate to recover from this property that  
\begin{eqnarray}
[N_1,R]=0
\end{eqnarray}
as it is clearly visible by inspection. The two types of supersymmetry that have been appearing in the literature are therefore in free fermion language essentially reduced to the more or less manifest symmetry-properties of the $R$-matrix $R \propto 1 - 2 N_1$ w.r.t the $\eta_i$ and $\eta^\dagger_i$ operators, $i=1,2$.

From a strict quantum-group viewpoint, we can interpret the appearance of the second supersymmetry as the presence of a Yangian symmetry at level $1$ with zero evaluation parameter $\hat{u}=0$ in Drinfeld's first realisation, which is rather typical of a conformal field theory \cite{Ale}. In fact, (\ref{lab}) can be read as 
\begin{eqnarray}
\Delta^{op} (\hat{q}) R = R \Delta(\hat{q}), \qquad \Delta(\hat{q}) = q \otimes E - E \otimes q =  e^{\frac{\theta_1}{2} + \theta_2} c_1 - e^{\frac{\theta_2}{2} + \theta_1} c_2, \qquad \hat{q}=0.
\end{eqnarray} 
By simply taking the Hermitian conjugate of (\ref{hermi}) we find
\begin{eqnarray}
\label{hermit}
(Q^\dagger_\pm)^{op} R = R \, Q^\dagger_\pm,
\end{eqnarray}
where {{}
\begin{eqnarray}
Q^\dagger_\pm =  \rho_{21,\pm}^* \, c_1^\dagger + \sigma_{21,\pm}^* \, c_2^\dagger, \qquad \rho_{21,\pm}^* = \sigma_{12,\pm}, \qquad \sigma_{21,\pm}^* = \rho_{12,\pm},
\end{eqnarray}}
reproducing again the string theory result \cite{JoakimAle}. In terms of the conformal field theory Yangian we have
\begin{eqnarray}
\Delta^{op} \big(\hat{q}^\dagger\big) R = R \, \Delta\big(\hat{q}^\dagger\big), \qquad \Delta\big(\hat{q}^\dagger\big) = q^\dagger \otimes E - E \otimes q^\dagger =  e^{\frac{\theta_1}{2} + \theta_2} c_1^\dagger - e^{\frac{\theta_2}{2} + \theta_1} c_2^\dagger, \qquad \hat{q}^\dagger=0.
\end{eqnarray}
We see that setting the evaluation parameter $\hat{u}=0$ is only one of the many possible choice, since adding any multiple of the level-$0$ supercharges by an even function of $\theta$ would still produce a symmetry. It is however the natural choice of organizing the symmetries according to the Yangian structure.

The fermionic number operator for one particle is given by
\begin{eqnarray}
b_0 = 1 - 2 c^\dagger c,  
\end{eqnarray}
and on two-particle states one has therefore the local coproduct
\begin{eqnarray}
{{} \Delta (b_0)} = 2 (1 - n_1 - n_2) = 2 (1 - N_1 - N_2).
\end{eqnarray}
From this observation it is immediate to see that, from $R \propto 1 - 2 N_1$, we have 
\begin{eqnarray}
[{{} \Delta (b_0)},R] = 0,
\end{eqnarray}
which is clear since the $AdS_3$ $R$-matrix preserves the fermionic number.

Another longer calculation reproduces also the following known result \cite{AntonioMartin}:
\begin{eqnarray}
{{} \Delta^{op} (b_1)} \, R = R \, {{} \Delta (b_1)},
\end{eqnarray} 
where the {\it secret symmetry} ${{} \Delta (b_1)}$ \cite{secret} is given in our new language by
\begin{eqnarray}
{{} \Delta (b_1)} = f(\theta) \big[n_1 + n_2\big] + c_1 c_2^\dagger + c_1^\dagger c_2,
\end{eqnarray}
with $f$ any even function of the variable $\theta = \theta_1 - \theta_2$ implementing the difference form.
In particular, since $n_1 + n_2$ is basically proportional to ${{} \Delta (b_0)}$ apart from a trivial shift, it is obvious that the first part is a symmetry, and we see that the core of the secret symmetry lies in the relation
\begin{eqnarray}
\{R, c_1 c_2^\dagger + c_1^\dagger c_2\}=0.
\end{eqnarray}
This can also be interpreted in terms of the conformal field theory Yangian \cite{Ale}, with the level $1$ secret symmetry being evaluated at $0$ corresponding to setting $f(\theta)=0$, which is again only one of the possible choices.

One final remark of this section is related to crossing symmetry. Despite the great simplification of using the free fermion variables, the operators $\eta$ and $\eta^\dagger$ do not appear to have nice transformation properties under crossing $\theta_1 \to \theta_1 + i \pi$. Therefore, one does not seem to learn much more than what is already known \cite{DiegoBogdanAle} in translating the crossing relations in the new language.

\subsection{$AdS_3$ with mixed flux}
\label{AdS3mixedf}
The $R$-matrix studied in \cite{gamma2} is another example which satisfies the free fermion condition: if we use the notations of \cite{gamma2}, we can see that
\begin{eqnarray}
a(\theta,k)^2 - b(\theta,k)^2 + c(\theta,k)=0,
\end{eqnarray}
which, considering that $d_{M} = \frac{1}{a(\theta,k)}$,  $a_{M} = - \frac{c(\theta,k)}{a(\theta,k)}$ and $c_{M} =b_{M}= i \frac{b(\theta,k)}{a(\theta,k)}$, exactly corresponds to $a_{M}d_{M} - b_{M}c_{M}=1$.  Here the index ${}_{M}$ refers again to the notation of \cite{MitevEtAl}, with the map to the notation of section \ref{sec:freefermion} given below (\ref{beloM}).

The transfer matrix $T_2$ is definitely more complicated than the previous case, and for explicit manipulations it can be efficiently treated by means of computer algebra. The change of variables engineered in this case is the same as in the previous section, with the only difference that we simply have now  
\begin{eqnarray}
\alpha = \frac{\pi}{4}.
\end{eqnarray}
We have not reproduced the procedure of appendix \ref{appB} this time and simply shown the result. With this transformation we obtain, with a suitable choice of normalisation of the $R$-matrix, that 
\begin{eqnarray}
T_2 = x + y N_1 + (y-2) N_2 + z \, N_1 \, N_2,
\end{eqnarray}
where\footnote{The case $k=1$ of these equations appears to be singular, but in fact this value should be treated separately as explained in \cite{gamma2}.}
\begin{eqnarray}
&&x = \frac{e^{-\frac{1}{2}(2 \theta_0 + \theta_1 + \theta_2)}\big(e^{2 i \frac{\pi}{k} + 2 \theta_0}-e^{\theta_1 + \theta_2}\big)}{e^{2 i \frac{\pi}{k}}-1},\nonumber\\
&&y = \frac{1 + e^{i \frac{\pi}{k}} - e^{i \pi \frac{\pi}{k} + \theta_0 - \frac{\theta_1}{2} - \frac{\theta_2}{2}}-e^{\frac{1}{2}(- 2 \theta_0 + \theta_1 + \theta_2)}}{1 + e^{i \frac{\pi}{k}}},\nonumber\\
&&z = 2 i \sinh \Big(\theta_0 - \frac{\theta_1}{2} - \frac{\theta_2}{2}\Big) \tan \frac{\pi}{2 k}.
\end{eqnarray}
As befitting the general formula (\ref{eq:Tdiag}), the same additive pattern we have described for the pure Ramond-Ramond case repeats here as well: the auxiliary Bethe equations are the same when written in terms of the function $b(\theta,k)$, and the eigenvalue of the transfer matrix for general $N$, when written in terms of the function $Y$ defined in \cite{gamma2}, has the exact same structure we observed in the previous section. This means that the ansatz (\ref{mani}), {\it mutatis mutandis}, works in exactly the same way. 

On of the specifics which is slightly different is the fact that already at $N=2$ we have a pair of finite solutions of the auxiliary Bethe equations, replacing the $\pm \infty$ of the pure Ramond-Ramond case. This means that the specific pattern of $\omega_i$ is different. In particular, we have at $N=2$ the two auxiliary roots 
\begin{eqnarray}
{{} e^{v_1}} = e^{- i \frac{\pi}{k}} e^{\frac{(\theta_1 + \theta_2)}{2}}, \qquad {{} e^{v_2}} = - {{} e^{v_1}}.
\end{eqnarray}

\subsection{The massive $AdS_3$ pure Ramond-Ramond case}
\label{massiveAdS3}
Let us see now how the general formalism applies to the massive $AdS_3$ pure Ramond-Ramond case. The free fermion condition is easily seen to hold equally well: one has in fact from \cite{CompleteT4} that the massive-sector $LL$ $R$-matrix reads
\begin{eqnarray}
R = A \, E_{11} \otimes E_{11} + B \, E_{11} \otimes E_{22} + C \, E_{21} \otimes E_{12} - F \, E_{22} \otimes E_{22} + D \, E_{22} \otimes E_{11} - E \, E_{12} \otimes E_{21},\nonumber
\end{eqnarray}  
where only for this particular subsection the symbols $A,B,C,D,E,F$ are used to denote the functions explicitly given in appendix M of \cite{CompleteT4}. One can directly verify that these functions satisfy the condition
\begin{eqnarray}
AF + BD = C^2, \qquad C=E,
\end{eqnarray}
which produce exactly the free fermion condition of \cite{MitevEtAl} or equivalently our (\ref{free}) restricted to 6-vertex models (see also comments in \cite{Borsato:2016xns}). Our formulas apply therefore, although the technical complication resides in the difficulty in getting further simplifications.

The same parameterisation we have been using for the massless case, with just a different choice of the parameter $\alpha$, works naturally for the massive case as well. We will again omit reproducing here the procedure of appendix \ref{appB}. Starting with the $R$-matrix, one can verify with computer algebra that the explicit value setting the $R$-matrix in free-fermion form is
\begin{eqnarray}
\label{tan}
\tan 2 \alpha = \frac{2 E}{D-B} = -2 \Big(\frac{x^-_p}{x^+_p}\Big)^{\frac{1}{4}}\Big(\frac{x^+_q}{x^-_q}\Big)^{\frac{1}{4}}\frac{\sqrt{x^-_p - x^+_p}\sqrt{x^-_q - x^+_q}}{\sqrt{\frac{x^-_p}{x^+_p}}(x^+_p - x^+_q) - \sqrt{\frac{x^+_q}{x^-_q}}(x^-_p - x^-_q)}.
\end{eqnarray}
It is interesting to note that the expression (\ref{tan}) behaves as a a $\frac{0}{0}$ in the BMN limit, and in fact tends to a finite non-trivial limit\footnote{As a consistency check, we have also verified that in massless right-right kinematics (\ref{tan}) tends to expression $\cot 2\alpha = \sinh \frac{\theta_{12}}{2}$ in (\ref{tend}) in the BMN limit.}. Most importantly, this limit is real, which might give some reassurance of a real angle $\alpha$ at least in a neighborhood of the physical region sufficiently close to the BMN point. We also notice that the expression for $R$ after the transformation characterised by (\ref{tan}) is purely in terms of $N_1$ and $N_2$, although the coefficients are rather complicated expressions of the representation variables. It is possible to see however that the coefficient of the term $N_1 N_2$ vanishes in the massless kinematics, which is another consistency check. 

Similarly, the two-site transfer matrix $T_2$ is expressed in free fermion form with the familiar canonical transformation. One can again check with computer algebra that the explicit value reads 
\begin{eqnarray}
\label{tann}
\tan 2 \alpha = \frac{2 E_{01} E_{02}}{A_{02}B_{01} - A_{01}B_{02} + D_{02}F_{01} - D_{01}F_{02}},
\end{eqnarray}
where we have indicated by the indices ${}_{01}$ and ${}_{02}$ the assignment of variables according to 
\begin{eqnarray}
T_2 = \mbox{str}_0 R_{01} R_{02}.
\end{eqnarray}
In particular, the complication arises when solving the auxiliary Bethe equations in terms of $x^\pm$ explicitly. Even for just one magnon, the expressions quickly become very cumbersome.

Besides being only dependent on $N_1$ and $N_2$, we have not been able to explicitly simplify $T_2$ much further, in fact even expression (\ref{tann}) remains rather difficult to reduce to small compact expressions in terms of the representation variables. We remain nevertheless firm in the knowledge that the formula we have derived for general 6-vertex models in section \ref{analternativetotheABA1} holds and, modulo the slight difficulty in containing the volume of the explicit expressions, achieves the complete diagonalisation and free fermion form for this case as well.  We can hope that solving the Bethe equations for the momentum carrying roots can help simplify the expressions further.

\section{8-vertex B model and application to $AdS_2/CFT_1$}
\label{8vmodel}
As opposed to 6-vertex models, we do not have a general formula displaying the free fermion form for 8-vertex models. We can in any case make progress and show that on two sites we can recast the transfer matrix and the Hamiltonian in the desired structure. This is particular relevant to the $AdS_2$ string theory models, which share the 8-vertex nature of the $R$-matrix. 

\subsection{Free oscillators\label{freeosc}}

The main new complication that arises for models of 8-vertex type is that there are terms that violate the fermion number. Because of this, the simple free fermion transformation that we used previously will not work and we need to add terms that mix $c$ and $c^\dag$. Thus, we will need to consider a general transformation of the form
\begin{align}
\binom{c}{c^\dag} = 
\begin{pmatrix}
U & V \\
V^* & U^*
\end{pmatrix}
 \binom{\eta}{\eta^\dag},
\end{align}
or in components
\begin{align}
& c_i=U_{ij}\,\eta_j + V_{ij}\,\eta_j^{\dagger},
&& c^\dag_i=U^*_{ij}\,\eta^\dag_j +V^*_{ij}\,\eta_{j},
\end{align}
\noindent
where sum in repeated indices is assumed.
In order to identify $c^\dag$ with the conjugate of $c$ we take $U^*$ to be the complex conjugate of $U$.
Imposing that the new oscillator basis $ \eta_i $ and $ \eta_i^\dagger $ also satisfies
\begin{align}
&\{\eta_i,\eta_j\}=0,   &&\{\eta_i^{\dagger},\eta_j^{\dagger}\}=0 && \{\eta_i,\eta_j^{\dagger}\}=\delta_{ij},
\end{align}
we find a set of consistency conditions on the matrices $U,V$. In particular, they need to satisfy
\begin{align}\label{eq:GenOsc}
&U V^T +VU^T = 0,
& U U^\dag + V V^\dag =1.
\end{align}
Notice that $U^* V^\dag +V^*U^\dag = 0$ as well. In case of the 6-vertex model we can set $V=0$ and the transformation is basically described by elements of $SU(N)$.

\subsection{8-vertex model and {\it pseudo - pseudo} vacuum}

The $R$-matrix for 8-vertex B presented in section \ref{sec:freefermion} can be written in terms of oscillators as
\begin{equation}
R(u,v)=m_1m_2r_1-n_1n_2r_4+m_1n_2r_2+n_1m_2r_3-(c_1c_2^{\dagger}-c_1^{\dagger}c_2)r_5-(c_1c_2+c_1^{\dagger}c_2^{\dagger})r_7
\end{equation}
where $ r_i\equiv r_i(u,v) $ and are given by \eqref{r1-8vB}-\eqref{r5678-8vB}\footnote{Notice that the calculations in this section were performed using the minus sign in equations \eqref{r1-8vB}-\eqref{r5678-8vB}, but the same could be done with the plus sign.}$ ^, $\footnote{Notice we already used here that $ r_6=r_5 $ and $ r_8=r_7 $ from \eqref{r5678-8vB}.}.

We then compute the monodromy for two sites 
\begin{equation}
T_2 = \mbox{str}_0 R_{01} R_{02}
\end{equation}
and obtain
%
\begin{align}
-T_2=~&
[r_1(\theta_0,\theta_1)r_1(\theta_0,\theta_2)-r_3(\theta_0,\theta_1)r_3(\theta_0,\theta_2)]\,(1-n_1-n_2 +n_1n_2)+\nonumber\\
& [r_1(\theta_0,\theta_2)r_2(\theta_0,\theta_1)+r_3(\theta_0,\theta_2)r_4(\theta_0,\theta_1)]\,n_1(1-n_2)+ \nonumber\\
&[r_1(\theta_0,\theta_1)r_2(\theta_0,\theta_2)+r_3(\theta_0,\theta_1)r_4(\theta_0,\theta_2)]\,n_2(1-n_2)+\nonumber\\
&[r_2(\theta_0,\theta_1)r_2(\theta_0,\theta_2)-r_4(\theta_0,\theta_1)r_4(\theta_0,\theta_2)]\,n_1n_2+\nonumber\\
&[r_5(\theta_0,\theta_2)r_7(\theta_0,\theta_1)-r_5(\theta_0,\theta_1)r_7(\theta_0,\theta_2)]\,(c_1c_2+c_1^{\dagger}c_2^{\dagger})+\nonumber\\
&[r_5(\theta_0,\theta_1)r_5(\theta_0,\theta_2)+r_7(\theta_0,\theta_1)r_7(\theta_0,\theta_2)]\,(c_1^{\dagger}c_2-c_1c_2^{\dagger}).
\label{T2for8vB}
\end{align}
For this simple case we can solve the conditions \eqref{eq:GenOsc} so that we have a canonical transformation explicitly and we find
\begin{align}
&V_{12}=-\frac{U_{11}V_{11}}{U_{12}}, && U^*_{11}=-\frac{ |U_{12}|^2}{U_{11}}\frac{|U_{12}|^2+|V_{11}|^2-1}{|U_{12}|^2+|V_{11}|^2}, \\
& V_{21}=\frac{U_{22}V_{11}}{U_{12}},  && U_{21}=\frac{U_{11}U_{22}}{U_{12}}\frac{|U_{12}|^2+|V_{11}|^2}{|U_{12}|^2+|V_{11}|^2-1}, \\
& V_{22}=-\frac{U_{21}V_{11}}{U_{12}}, && U^*_{22}=-\frac{|U_{12}|^2}{U_{22}}\frac{|U_{12}|^2+|V_{11}|^2-1}{|U_{12}|^2+|V_{11}|^2} ,
\label{anticomutccdagger}
\end{align}
together with their starred versions.

We then proceed to fix the remaining $ U_{ij} $ and $ V_{ij} $ by diagonalizing $ T_2 $ in equation \eqref{T2for8vB}. Requiring that the nondiagonal terms vanish yields the following constraints
\begin{align}
U_{11}=&-i\frac{\sinh(\mu_2)}{2} \frac{U_{12}}{U_{22}V_{11}}\frac{\left(|U_{12}|^2-|V_{11}|^2\right)\left(|U_{12}|^2+|V_{11}|^2-1\right)}{|U_{12}|^2+|V_{11}|^2}
\label{eqA11},\\
U_{22}=&\frac{\sinh(\mu_1)}{2} \frac{2|U_{12}|^2+2|V_{11}|^2-1}{|U_{12}|^2+|V_{11}|^2}U_{12},
\label{eqA22}
\end{align}
with 
\begin{align}
& \frac{\sinh(\mu_2)}{2}=i\frac{r_5(\theta_0,\theta_2)r_7(\theta_0,\theta_1)-r_5(\theta_0,\theta_1)r_7(\theta_0,\theta_2)}{r_1(\theta_0,\theta_1)r_1(\theta_0,\theta_2)-r_2(\theta_0,\theta_1)r_2(\theta_0,\theta_2)-r_3(\theta_0,\theta_1)r_3(\theta_0,\theta_2)+r_4(\theta_0,\theta_1)r_4(\theta_0,\theta_2)},\nonumber\\
&\frac{\sinh(\mu_1)}{2}=\frac{r_5(\theta_0,\theta_1)r_5(\theta_0,\theta_2)+r_7(\theta_0,\theta_1)r_7(\theta_0,\theta_2)}{r_1(\theta_0,\theta_2)r_2(\theta_0,\theta_1)-r_1(\theta_0,\theta_1)r_2(\theta_0,\theta_2)+r_3(\theta_0,\theta_2)r_4(\theta_0,\theta_1)-r_3(\theta_0,\theta_1)r_4(\theta_0,\theta_2)}.
\label{mudefinitions}
\end{align}
Additionally we have 
\begin{align}
& U_{11}^2=\frac{|U_{12}|^2V^*_{11}}{V_{11}}\frac{|U_{12}|^2+|V_{11}|^2-1}{|U_{12}|^2+|V_{11}|^2},
&& U_{22}^2=-U_{12}^2\frac{|U_{12}|^2+|V_{11}|^2-1}{|U_{12}|^2+|V_{11}|^2}.\label{eqA22squared}
\end{align}
In order to make equations \eqref{eqA11} and \eqref{eqA22} consistent with \eqref{eqA22squared} and complex conjugation we need to satisfy the following conditions
\begin{align}
& |V_{11}|^2 =\frac{1}{4}\text{sech}\mu_1\text{sech}\mu_2\sinh^2\left(\frac{\mu_2}{2}\right)\left(e^{\frac{\mu_1}{2}}\mp e^{-\frac{\mu_1}{2}}\right)^2,\label{X}\\
& |U_{12}|^2 =\frac{1}{4}\text{sech}\mu_1\text{sech}\mu_2\cosh^2\left(\frac{\mu_2}{2}\right)\left(e^{\frac{\mu_1}{2}}\mp e^{-\frac{\mu_1}{2}}\right)^2\label{Y}.
\end{align}
%
Using all the definitions above one finally arrives at the diagonal $ T_2 $
\begin{equation}
T_2=t_1 1+t_2 N_1+t_3 N_2+t_4 N_1N_2
\label{diagonalizedT2}
\end{equation}
where $ N_i=\eta^{\dagger}_i\eta_i $, $ t_i\equiv t_i(\theta_0,\theta_1,\theta_2)$ and
\begin{align}
t_1=& -\frac{|V_{11}|^2}{|V_{11}|^2-|U_{12}|^2}\left(r_2(\theta_0,\theta_1)r_2(\theta_0,\theta_2)-r_4(\theta_0,\theta_1)r_4(\theta_0,\theta_2)\right)\nonumber\\
& +\frac{|U_{12}|^2}{|V_{11}|^2+|U_{12}|^2}\left(r_1(\theta_0,\theta_1)r_1(\theta_0,\theta_2)-r_3(\theta_0,\theta_1)r_3(\theta_0,\theta_2)\right),\label{t1}
\end{align}
\begin{align}
t_2=&-\frac{|V_{11}|^2+|U_{12}|^2-1}{2(|V_{11}|^2+|U_{12}|^2)-1}\left(r_1(\theta_0,\theta_1)r_2(\theta_0,\theta_2)+r_3(\theta_0,\theta_1)r_4(\theta_0,\theta_2)\right)\nonumber\\
&+\frac{|V_{11}|^2}{|V_{11}|^2+|U_{12}|^2}\left(r_2(\theta_0,\theta_1)r_2(\theta_0,\theta_2)-r_4(\theta_0,\theta_1)r_4(\theta_0,\theta_2)\right)\nonumber\\
&-\frac{|V_{11}|^2+|U_{12}|^2}{2(|V_{11}|^2+|U_{12}|^2)-1}\left(r_1(\theta_0,\theta_2)r_2(\theta_0,\theta_1)+r_3(\theta_0,\theta_2)r_4(\theta_0,\theta_1)\right)\nonumber\\
&-\frac{|U_{12}|^2}{|V_{11}|^2+|U_{12}|^2}\left(r_1(\theta_0,\theta_1)r_1(\theta_0,\theta_2)-r_3(\theta_0,\theta_1)r_3(\theta_0,\theta_2)\right),\label{t2}
\end{align}
\begin{align} 
t_3=&-\frac{|V_{11}|^2+|U_{12}|^2}{2(|V_{11}|^2+|U_{12}|^2)-1}\left(r_1(\theta_0,\theta_1)r_2(\theta_0,\theta_2)+r_3(\theta_0,\theta_1)r_4(\theta_0,\theta_2)\right)\nonumber\\
&+\frac{|V_{11}|^2}{|V_{11}|^2+|U_{12}|^2}\left(r_2(\theta_0,\theta_1)r_2(\theta_0,\theta_2)-r_4(\theta_0,\theta_1)r_4(\theta_0,\theta_2)\right)\nonumber\\
&-\frac{|V_{11}|^2+|U_{12}|^2-1}{2(|V_{11}|^2+|U_{12}|^2)-1}\left(r_1(\theta_0,\theta_2)r_2(\theta_0,\theta_1)+r_3(\theta_0,\theta_2)r_4(\theta_0,\theta_1)\right)\nonumber\\
&-\frac{|U_{12}|^2}{|V_{11}|^2+|U_{12}|^2}\left(r_1(\theta_0,\theta_1)r_1(\theta_0,\theta_2)-r_3(\theta_0,\theta_1)r_3(\theta_0,\theta_2)\right),\label{t3}
\end{align}
\begin{align}
t_4=& -\left(r_1(\theta_0,\theta_1)-r_2(\theta_0,\theta_1)\right)\left(r_1(\theta_0,\theta_2)-r_2(\theta_0,\theta_2)\right)\nonumber\\
&+\left(r_3(\theta_0,\theta_1)+r_4(\theta_0,\theta_1)\right)\left(r_3(\theta_0,\theta_2)+r_4(\theta_0,\theta_2)\right).
\label{t4}
\end{align}

Remarkably, we can factorize $ T_2 $  as

\begin{equation}
T_2=t_1\left(1+\frac{t_2}{t_1}N_1\right)\left(1+\frac{t_3}{t_1}N_2\right).
\end{equation}
In order to this factorization be correct we need that $ t_i $ as in \eqref{t1}-\eqref{t4} satisfy $  t_2t_3=t_1t_4 $. We checked this property numerically for $ r_i $ as in \eqref{r1-8vB}-\eqref{r5678-8vB}  with both complex and real values of $ \{k, \theta_0, \theta_1, \theta_2 \} $.

At this point we notice that, although these models do not possess a pseudovacuum, we can find, thanks to the particle-hole transformation we have performed, a state which behaves in a similar way, and allows us to construct the spectrum in the usual fashion a genuine pseudovacuum would allow us to do. We call this state the \textit{pseudo pseudo} vacuum\footnote{While we have found it for $N=2$, we do not know whether it would actually exist for higher values of $N$.}. It is given by 

\begin{equation}
| \text{vac} \rangle = \kappa \bigg[ |\psi\rangle \otimes |\psi\rangle + i\, \coth \left(\frac{\mu_2}{2}\right)|\phi\rangle \otimes |\phi\rangle \bigg],
\label{vacuum}
\end{equation} 
such that $ \eta_1 | \text{vac} \rangle =0=\eta_2 | \text{vac} \rangle$ and $\kappa $ is the normalisation factor. 

The eigenvalues $ \lambda_i $ of the transfer matrix are given by 

\begin{itemize}
	\item $ \lambda_1 =t_1 $ for the vacuum $ |\text{vac}\rangle $;
	\item $ \lambda_2=t_1+t_2 $ for the eigenvector $\eta_1^\dagger|\text{vac}\rangle $;
	\item $ \lambda_3=t_1+t_3 $ for the eigenvector $\eta_2^\dagger|\text{vac}\rangle $;
	\item and $ \lambda_4=t_1+t_2+t_3+t_4 $ for the eigenvector $\eta_1^\dagger\eta_2^\dagger|\text{vac}\rangle $
\end{itemize}
where $ t_i $ are given in equations \eqref{t1}-\eqref{t4}.

\subsection{The $AdS_2$ transfer matrix}
\label{section53}
We shall now apply this formalism to the $R$-matrix dubbed {\it Solution 3} in \cite{Andrea2}, corresponding to a particular situation in massless relativistic $AdS_2$ integrable superstrings. This $R$-matrix still satisfies the free fermion condition, as was noticed in \cite{Andrea2,Ale}. This fact was in fact crucial for the purposes of \cite{Andrea2,Ale}, since it was only thanks to this condition that a procedure devised in \cite{Felderhof} and used in \cite{Ahn} for ${\cal{N}}=1$ supersymmetric Sine-Gordon could be repeated in the $AdS_2$ case. The $AdS_2$ transfer matrix shares with those other models the same feature of the absence of a pseudo-vacuum state for the algebraic Bethe Ansatz, invalidating its applicability. The method which \cite{Andrea2} used was instead to obtain the auxiliary Bethe equations via Zamolodchikov's inversion relations, and then \cite{Ale} brute-force computed the transfer-matrix eigenvalues for a few number of sites and extrapolated the expression for the eigenvalues at any $N$. This is only viable thanks to the free fermion condition, which allows the very first step to be implemented.

With an appropriate normalisation of the $R$-matrix (and again, as always in this paper, neglecting for these purposes the overall dressing factor), the transfer matrix $T_2$, which has been computed in \cite{Ale}, translates here into
\begin{eqnarray}
T_2 = 1 - n_1 - n_2 + 2 n_1 n_2 + g_{01} g_{02} (c_1 \, c_2 - c_1^\dagger c_2^\dagger), \qquad g(\theta) = e^{-\frac{\theta}{2}}.
\end{eqnarray}
As we can see, this case is akin to an $8$-vertex model due to the non conservation of the fermion number. The canonical transformation reduced from section \ref{freeosc} is
\begin{eqnarray}
c_1 = \cos \alpha \, \eta_1 - \sin \alpha \, \eta^\dagger_2, \qquad c_2^\dagger = \sin \alpha \, \eta_1 + \cos \alpha \, \eta^\dagger_2.
\end{eqnarray} 
with, 
\begin{eqnarray}
\alpha = \frac{\pi}{4}
\end{eqnarray} 
which ultimately allows the rewriting
\begin{eqnarray}
T_2 = (1+ g_{01}g_{02})\big(1 - N_2 - N_1\big) + 2 N_1 N_2, \qquad N_i = \eta^\dagger_i \eta_i. 
\end{eqnarray}

{{} As we have anticipated}, it can be verified that there is a state annihilated by both $\eta_1$ and $\eta_2$, which will effectively play the substitute of the pseudovacuum on two physical spaces. The absence of global pseudovacua for all $N$ forced us to employ alternative techniques to the Algebraic Bethe Ansatz in \cite{Andrea2,Ale}, however here we saw that there is a way to simulate the existence of a pseudovacuum for $N=2$.   
{{} This} pseudo - pseudo vacuum is given by 
\begin{eqnarray}
|\mbox{vac}\rangle = \frac{1}{\sqrt{2}} \big(|0\rangle - c_1^\dagger c_2^\dagger |0\rangle\big) = \frac{1}{\sqrt{2}} \big(|\phi \rangle \otimes |\phi\rangle - |\psi\rangle \otimes |\psi\rangle\big), \qquad \eta_1 |\mbox{vac}\rangle = 0 = \eta_2 |\mbox{vac}\rangle.
\end{eqnarray}
The spectrum is then populated in the usual fashion. 

The paper \cite{Ale} extended the derivation of the auxiliary Bethe equations via the method of inversion relations to the {\it massive} $AdS_2$ $R$-matrix as well. It is now no surprise that this could be achieved (although not much further progress could be made due to technical complication of the $R$-matrix entries) once again because of the general result we have proven {{} in section \ref{analternativetotheABA}}. The massive $AdS_2$ $R$-matrix satisfies equally well the free fermion condition following those very general arguments.  

One peculiar difference between the massive and the massless $AdS_2$ case is that $R(p,p')$ equals the graded permutation only for the massive $R$-matrix, not the massless one. This is due to a non-commutativity of limits (equal arguments {\it vs} massless limit). The massless limit for the $AdS_2$ case is especially subtle and it has been discussed in detail in \cite{Hoare:2014kma}. As a consequence, the massive $R$-matrix sits inside the classification of \cite{MCAAP} (and admits in principle a nearest-neighbour Hamiltonian at equal inhomogeneities), while the massless one does not. It is still true that the same form of the massless $R$-matrix is valid for the BMN limit as well as for the full non-relativistic case, with the same change of variable as in $AdS_3$, as shown in \cite{gamma1}. It is also still true that the massless $R$-matrix equally satisfies the free fermion condition. 

\section{Free fermion condition for $ AdS_5 $ sector}\label{sec:16}

In this section we will finally apply our approach discussed in Section \ref{sec:freefermion} 	to obtain a free fermion condition to the case of a 4-dimensional Hilbert space. We will focus on the class of model whose Hamiltonian and $R$-matrix exhibit $\mathfrak{su}(2)\oplus \mathfrak{su}(2)$ symmetry \cite{MCAAP2,MAAP}. This class of models is particularly important because it contains the $R$-matrices of the {{} $AdS_5 \times S^5$} superstring sigma model and the one-dimensional Hubbard model.

In order to achieve this, we follow the procedure outlined in Section \ref{suth}. In matrix form, the Hamiltonian is given by
\setcounter{MaxMatrixCols}{20}
\begin{align}
\mathcal{H}_{12}^{(mat)}=\tiny{\begin{pmatrix}
h_1+h_2 & 0 & 0 & 0 & 0 & 0 & 0 & 0 & 0 & 0 & 0 & 0 & 0 & 0 & 0 & 0 \\
 0 & h_1 & 0 & 0 & h_2 & 0 & 0 & 0 & 0 & 0 & 0 & h_{10} & 0 & 0 & -h_{10} & 0 \\
 0 & 0 & h_4 & 0 & 0 & 0 & 0 & 0 & h_7 & 0 & 0 & 0 & 0 & 0 & 0 & 0 \\
 0 & 0 & 0 & h_4 & 0 & 0 & 0 & 0 & 0 & 0 & 0 & 0 & h_7 & 0 & 0 & 0 \\
 0 & h_2 & 0 & 0 & h_1 & 0 & 0 & 0 & 0 & 0 & 0 & -h_{10} & 0 & 0 & h_{10} & 0 \\
 0 & 0 & 0 & 0 & 0 & h_1+h_2 & 0 & 0 & 0 & 0 & 0 & 0 & 0 & 0 & 0 & 0 \\
 0 & 0 & 0 & 0 & 0 & 0 & h_4 & 0 & 0 & h_7 & 0 & 0 & 0 & 0 & 0 & 0 \\
 0 & 0 & 0 & 0 & 0 & 0 & 0 & h_4 & 0 & 0 & 0 & 0 & 0 & h_7 & 0 & 0 \\
 0 & 0 & h_5 & 0 & 0 & 0 & 0 & 0 & h_6 & 0 & 0 & 0 & 0 & 0 & 0 & 0 \\
 0 & 0 & 0 & 0 & 0 & 0 & h_5 & 0 & 0 & h_6 & 0 & 0 & 0 & 0 & 0 & 0 \\
 0 & 0 & 0 & 0 & 0 & 0 & 0 & 0 & 0 & 0 & h_8+h_9 & 0 & 0 & 0 & 0 & 0 \\
 0 & h_3 & 0 & 0 & -h_3 & 0 & 0 & 0 & 0 & 0 & 0 & h_8 & 0 & 0 & h_9 & 0 \\
 0 & 0 & 0 & h_5 & 0 & 0 & 0 & 0 & 0 & 0 & 0 & 0 & h_6 & 0 & 0 & 0 \\
 0 & 0 & 0 & 0 & 0 & 0 & 0 & h_5 & 0 & 0 & 0 & 0 & 0 & h_6 & 0 & 0 \\
 0 & -h_3 & 0 & 0 & h_3 & 0 & 0 & 0 & 0 & 0 & 0 & h_9 & 0 & 0 & h_8 & 0 \\
 0 & 0 & 0 & 0 & 0 & 0 & 0 & 0 & 0 & 0 & 0 & 0 & 0 & 0 & 0 & h_8+h_9
\end{pmatrix}},
\end{align}
having here as in other instances omitted the parametric dependence of the $h_i$s for ease of notation. The $R$-matrix takes the same form, but with coefficients $r_i(u,v)$.

As explained in Section \ref{suth}, we should first substitute the Hamiltonian and the $R$-matrix in the Sutherland equations \eqref{Sutherland}. Then, we solved them for the derivatives $\dot r_i$ and $r_i'$. This was done formally considering the derivatives $\dot r_i$ and $r_i'$ as independent variables from the  $r_i$. It is worth mentioning that in order to avoid divergences it is important to properly chose the order of solutions of the equations. In our case, we used the list of models in \cite{MCAAP2} as test. In this way we were able to keep the divergences under control and to chose the correct branch of solutions\footnote{Some equations in fact can be factorised in the form $f g =0$, with $f$ and $g$ functions of the $r_i$s. We used the test to select whether the solution was $f=0$ or $g=0$. If one chose the models classified in \cite{MAAP} as test, some choices would have been different.}.
In particular, we find the following very simple conditions
\begin{align}
&h_3(v) r_{10}r_5=h_{10}(v) r_3 r_7,\\
&h_3(u) r_{10} r_7=h_{10}(u) r_3 r_5.
\end{align}
At this point, we need to distinguish two cases $r_3\neq 0$ and $r_3=0$. In the next subsection the physical meaning of this choice will be clear.

\subsection{Case $r_3\neq 0$}
\label{r3no0}
The one-dimensional Hubbard model and the $S$-matrix of $AdS_5 \times S^5$ superstring sigma model fall in this class of models. Similar to the Baxter condition, we find
\begin{align}
& \frac{r_4^2-r_1(r_1+r_2)}{r_3r_7}=\frac{h_2(v)}{h_3(v)},\\
& \frac{r_6^2-r_8(r_8+r_9)}{r_3 r_7}=\frac{h_9(v)}{h_{3}(v)},\\
& \frac{r_6^2-r_1(r_1+r_2)}{r_3r_5}=\frac{h_2(u)}{h_3(u)},\\
& \frac{r_4^2-r_8(r_8+r_9)}{r_3 r_5}=\frac{h_9(u)}{h_{3}(u)}.
\end{align}
We remark that $h_3=0$ does not cause divergences, since $r_3\neq 0$ implies that $h_3$ is also non-zero. Furthermore in all the models from \cite{MCAAP2}  that fall into this category $r_5, r_7\ne 0$. After solving some of the first Sutherland equations \eqref{Sutherland} for the $h$s and plugging the solutions into the others, we are left with four conditions
\begin{align}
&r_4 r_6+r_1 r_8=r_3 r_{10},\label{mitev1}\\
&r_5 r_7-r_4 r_6=\left(r_1+r_2\right) \left(r_8+r_9\right),\label{mitev2}\\
&r_3 r_{10}+r_5 r_7=r_2 r_9,\label{mitev3}\\
&\left(r_1+r_8\right) \left(r_1+r_2+r_8+r_9\right)=\left(r_4-r_6\right)^2 \label{nomitev4}.
\end{align}
It can be shown that these four conditions could have been similarly derived from the second Sutherland equations. We can notice that \eqref{mitev1}-\eqref{mitev3} were obtained also in \cite{MitevEtAl}. However, we find the \textit{additional} condition \eqref{nomitev4}.

By using the regularity condition $R(u,u)=P$ and from the definition of the Hamiltonian, we can derive that \eqref{mitev1}-\eqref{nomitev4} impose some very simple constraints on the Hamiltonian\footnote{\eqref{condHam1} and \eqref{condHam2} were derived differentiating \eqref{mitev3} and \eqref{nomitev4}, \eqref{condHam3} differentiating  \eqref{mitev1} twice.}  
\begin{align}
&h_1+h_8=h_4+h_6,\label{condHam1}\\
&h_2=-h_9,\label{condHam2}\\
&h_5 h_7+h_2 h_9= h_3 h_{10}\label{condHam3}.
\end{align}
Furthermore, differentiating two times and combining \eqref{mitev1}-\eqref{nomitev4} in order to get rid of the second derivatives, we also got
$$\left(h_5+h_7\right)^2=\left(h_2-h_9\right) \left(h_1+h_2-h_8-h_9\right).$$

\subsection{ Case $r_3=0$}

One can check that the case $r_3=0$ and $r_{10}\neq 0$ 
actually satisfies \eqref{mitev1}-\eqref{nomitev4} as well. Hence, we can restrict to the case where $r_{10}=0$. Analyzing this case leads to a set of factorised equations
\begin{align}
&\frac{r_1}{r_6}=\frac{h_2(v)}{h_7(v)},
&&\frac{r_8}{r_4}=\frac{h_9(v)}{h_5(v)},
&&\frac{r_1 r_5 r_7}{r_2^2r_4}=\frac{h_2(v)}{h_5(v)},\\
&\frac{r_1}{r_4}=\frac{h_2(u)}{h_5(u)},
&&\frac{r_8}{r_6}=\frac{h_9(u)}{h_7(u)}, 
&&\frac{r_1 r_5 r_7}{r_2^2r_6}=\frac{h_2(u)}{h_7(u)}.
\end{align}
Plugging the solutions for the $h$s into the Sutherland equations, it can be noticed that it is not possible to find conditions that hold in all the cases. One should consider four possible subcases separately: $r_1\neq 0 \, r_8\neq 0$, $r_1\neq 0\, r_8= 0$, $r_1=0\, r_8\neq 0$ and $r_1= 0\, r_8=0$. The numbers of subcases can be reduced using the transformations on the $R$-matrix that preserve integrability, \cite{MCAAP,MCAAP2,MAAP}. In particular, we will use
\begin{itemize}
\item[1] Local basis transformations: $R_{12}\to W_1(u) W_2(v) R_{12} (W_1(u) W_2(v))^{-1}$,
\item[2] Twist: given $[W_1(u)W_2(v),R_{12}(u,v)]=0,$ $R_{12}\to W_2(u)R_{12} W_1^{-1}(v).$
\end{itemize}
The subcases to be considered are then

\subsubsection*{Subcase $r_1\neq 0, r_8\neq 0$}
We obtain
\begin{align}
&r_5 {r_7}=r_2 r_9,\\
&r_4 {r_6} r_9=r_2 r_8^2,\\
&r_1^2 r_9^2=r_2^2 r_8^2.
\end{align}
These imply \eqref{condHam1} on the entries of the Hamiltonian together with $h_2^2= h_9^2$, $h_5^2 h_7^2 = h_9^2$.

\subsubsection*{Subcase $r_1\neq0, r_8=0$}
In this case, the entries of the $R$-matrix satisfy
\begin{align}
&r_2^2 r_4 r_6=r_1^2 r_5 r_7,\\
&r_4 r_5 r_6 r_7=\left(r_5 r_7-r_2 r_9\right){}^2.
\end{align}
And on the Hamiltonian
\begin{align}
& h_5 h_7= h_2^2,\\
&  h_5 h_7= \left(h_1-h_4-h_6+h_8\right){}^2.
\end{align}
We can notice that the subcase $r_1=0, r_8\neq0$ can be recovered from $r_1\neq0, r_8=0$ by performing an off-diagonal basis transformation\footnote{The action of the off-diagonal basis transformation needed is to swap $g_1 \leftrightarrow g_8, g_3 \leftrightarrow g_{10}, g_2 \leftrightarrow g_9, g_1 \leftrightarrow g_8, g_4 \leftrightarrow g_6, g_5 \leftrightarrow g_7,$ where $g$ is either $h$ or $r$.} on the $R$-matrix and on the Hamiltonian.

\subsubsection*{Subcase $r_1=0, r_8= 0$}
There are two {{} possibilities. If}
\begin{align}
&h_5(v) r_6 r_9=h_7(v) r_2 r_4 \label{caser10r80} 
&&\text{and}
&& h_7(u) r_4 r_9=h_5(u) r_2 r_6
\end{align}
there are no additional conditions on the entries of the $R$-matrix.
If \eqref{caser10r80} are not verified, we obtain the condition
\begin{align}
&r_4 r_6-r_5 r_7+r_2 r_9=0
\end{align}
that implies on the Hamiltonian \eqref{condHam1}.

In the following section we would like to show how to rewrite the Hamiltonian to make the free fermion nature explicit.

\subsection{Towards a free fermion Hamiltonian}

Here we will mainly follow section \ref{analternativetotheABA} and \cite{Lieb,Lieb2}. As already mentioned, the Hilbert space is four dimensional and is spanned by two bosons $|\phi_{1,2}\rangle$ and two fermions $|\psi_{1,2}\rangle$. We introduce two sets of canonical fermionic creation and annihilation operators
$c^\dagger_{\alpha,j},c_{\alpha,j}$ where $\alpha =\, \uparrow,\downarrow$ is the spin and  $j$ is the site of the chain (running from $1$ to {{} the chain length $N$}). If we denote the vacuum by $|0\rangle$ such that $c_{\alpha,j}|0\rangle=0$, then our local Hilbert space is spanned by
\begin{align}
&|\phi_1\rangle= |0\rangle,
&&|\phi_2\rangle= c^\dagger_{\uparrow,j} c^\dagger_{\downarrow,j} |0\rangle,
&&|\psi_1\rangle = c^\dagger_{\uparrow,j} |0\rangle,
&&|\psi_2\rangle = c^\dagger_{\downarrow,j} |0\rangle.
\label{phi12psi12}
\end{align}
Those oscillators satisfy the usual anti-commutation relations
\begin{align}\label{eq:oscdef}
&\{c^\dagger_{\alpha,i},c_{\beta,j} \} = \delta_{\alpha\beta}\delta_{ij},
&&\{c_{\alpha,i},c_{\beta,j} \} = 0,
&&\{c^\dagger_{\alpha,i},c^\dagger_{\beta,j} \} = 0,
\end{align}
where $\alpha$ and $\beta$ can be either $\uparrow$ and $\downarrow$.
The $\mathfrak{su}(2)\oplus \mathfrak{su}(2)$ $R$-matrix is given by
\begin{align}
R |\phi_a \phi_b \rangle &= r_1  |\phi_a \phi_b \rangle + r_2  |\phi_b \phi_a \rangle + r_3 \epsilon_{ab}\epsilon_{\alpha\beta} |\psi_\alpha \psi_\beta \rangle, \label{eq:Rphiphi}\\
R |\phi_a \psi_\beta \rangle &= r_4   |\phi_a \psi_\beta \rangle + r_5  |\psi_\beta \phi_a \rangle, \label{Rphipsi} \\
R |\psi_\alpha \phi_b \rangle &= r_6  |\psi_\alpha \phi_b \rangle + r_{7}  |\phi_b \psi_\alpha \rangle,  \label{Rpsiphi}\\
R|\psi_\alpha \psi_\beta \rangle &= r_8  |\psi_\alpha \psi_\beta \rangle + r_9  |\psi_\beta \psi_\alpha \rangle + r_{10} \epsilon_{ab}\epsilon_{\alpha\beta} |\phi_a \phi_b \rangle\label{Rpsipsi},
\end{align}
where $r_i\equiv r_i(u,v)$. The $R$-matrix can be written completely in term of oscillators 
\begin{align}
R_{12}^{(osc)} =
&~\sum_{\substack{\{\alpha,\beta\}=\{\uparrow,\downarrow\}, \\ \{\downarrow,\uparrow\}}} \Big[ ( c^\dagger_{\alpha,1}c_{\alpha,2} + c_{\alpha,1} c^\dagger_{\alpha,2})(C_1 + C_2 (n_{\beta,1} - n_{\beta,2})^2 ) + \nonumber\\
&\qquad ( c^\dagger_{\alpha,1}c_{\alpha,2} - c_{\alpha,1} c^\dagger_{\alpha,2})(C_3(n_{\beta,1}-\frac{1}{2}) + C_4 (n_{\beta,2}-\frac{1}{2}) )\Big]\nonumber \\
&~+( c^\dagger_{\uparrow,1}c^\dagger_{\downarrow,1}c_{\uparrow,2}c_{\downarrow,2} +  c_{\uparrow,1}c_{\downarrow,1}c^\dagger_{\uparrow,2}c^\dagger_{\downarrow,2}) C_5 + \nonumber
( c^\dagger_{\uparrow,1}c_{\downarrow,1}c^\dagger_{\downarrow,2}c_{\uparrow,2} +  c^\dagger_{\downarrow,1}c_{\uparrow,1}c^\dagger_{\uparrow,2}c_{\downarrow,2}) C_6 \nonumber\\
&~ + C_7 (n_{\uparrow,1}-\frac{1}{2})(n_{\downarrow,1}-\frac{1}{2}) + C_8 (n_{\uparrow,2}-\frac{1}{2})(n_{\downarrow,2}-\frac{1}{2})\nonumber  \\
&~ + C_9 (n_{\uparrow,1}-n_{\downarrow,1})^2 (n_{\uparrow,2}-n_{\downarrow,2})^2 +\nonumber\\
&~ + (C_5-C_6) (n_{\uparrow,1}n_{\downarrow,1}+n_{\uparrow,2}n_{\downarrow,2}-1)(n_{\uparrow,1}-n_{\uparrow,2})(n_{\downarrow,1}-n_{\downarrow,2})\nonumber \\ 
&~ +\frac{1}{2} C_5 ((n_{\uparrow,1}-n_{\downarrow,2})^2+(n_{\downarrow,1}-n_{\uparrow,2})^2)+C_0,
\end{align}
\noindent
where $n_{\alpha,k}\equiv c^\dagger_{\alpha,k} c_{\alpha,k}$, $\,\,\,C_0,\dots,C_9$ are ten functions dependent on the {{} parameters} $(u,v)$ and related to the $r_i$ in the following way
\begin{align}
& C_0=\frac{1}{2} \left(\left(r_4+r_6\right) \tau +r_2\right),\,C_1=\frac{1}{2} \left(r_7-r_5\right),\,C_2=\frac{1}{2} \left(r_5-r_7-\sigma (r_3+r_{10})\right), \nonumber\\
& C_3=\frac{1}{2} \left( \sigma (r_3-r_{10} ) +r_5+r_7\right),\,C_4=\frac{1}{2} \left(r_5+r_7-\sigma (r_3 -r_{10})\right),\,C_5=-r_2,\,C_6=-r_9,\nonumber\\
& C_7=-2 r_6 \tau +2 r_1+r_2,\, C_8=-2 r_4 \tau +2 r_1+r_2,\,C_9=-(r_4+r_6) \tau +r_1+r_2-r_8-r_9,
\end{align}
$\sigma$ and $\tau$ are arbitrary signs\footnote{Interestingly, using oscillators, the arbitrariness of $\sigma$ and $\tau$ naturally emerges. Those can be understood as a local basis transformation ($\sigma$) and a twist ($\tau$).}.
By taking the logarithmic derivative of the $R$-matrix we get the following expression for the Hamiltonian
\begin{align}
\mathcal{H}_{12}^{(osc)} &= (h_2+h_9) \left[\sum _{\alpha =\{\uparrow ,\downarrow \}} n_{\alpha ,2}  n_{\alpha ,1}-\left(n_{\uparrow ,1}+n_{\downarrow ,1}\right)  n_{\uparrow ,2}  n_{\downarrow ,2}-\left(n_{\uparrow ,2}+n_{\downarrow ,2}\right)  n_{\uparrow ,1}  n_{\downarrow ,1} + ~2n_{\downarrow ,1}  n_{\uparrow ,1}  n_{\downarrow ,2}  n_{\uparrow ,2} \right]\nonumber\\
&~+(h_1-h_4-h_6+h_8)\Big[ \left(n_{\uparrow ,2}+n_{\downarrow ,2}\right)  \left(n_{\uparrow ,1}+n_{\downarrow ,1}\right)-2\left(n_{\uparrow ,1}+n_{\downarrow ,1}\right)  n_{\downarrow ,2}  n_{\uparrow ,2}\nonumber\\
&-2 \left(n_{\uparrow ,2}+n_{\downarrow ,2}\right) n_{\downarrow ,1}  n_{\uparrow ,1} + 4n_{\downarrow ,1}  n_{\uparrow ,1}  n_{\downarrow ,2}  n_{\uparrow ,2} \Big] + \mathcal{H}_{12}^{(1pt)}+\mathcal{H}_{12}^{(2pt)} - \left[ (h_5+h_7)\tau+(h_{10}-h_3)\sigma \right]\mathcal{H}_{12}^{(3pt)},
\label{eq:ham1616}
\end{align}
where $\mathcal{H}_{12}^{(k\, pt)}$ identifies the subsector of $k$-particles. This is possible since
\begin{align}
&[n_{TOT},\mathcal{H}_{12}^{(osc)}]=0, && n_{TOT}=\sum _{\alpha =\{\uparrow ,\downarrow \}} n_{\alpha,1}+n_{\alpha,2},
\end{align}
so the total number of particles is conserved. The $ k $-particle subsectors are represented by
\begin{equation}\label{eq:n-particle}
	\begin{aligned}
		\mathcal{H}_{12}^{(1pt)} &= \left(\left( c^\dagger_{\uparrow ,1} c_{\uparrow ,2}+ c^\dagger_{\downarrow ,1} c_{\downarrow ,2}\right)h_5+\left( c^\dagger_{\uparrow ,2} c_{\uparrow ,1}+ c^\dagger_{\downarrow ,2} c_{\downarrow ,1}\right)h_7\right)\tau+h_6\left(n_{\uparrow ,1}+n_{\downarrow ,1}\right)\\
		& +h_4\left(n_{\uparrow ,2}+n_{\downarrow ,2}\right) - (h_1+h_2)\left(-1+n_{TOT}\right),\\
		\mathcal{H}_{12}^{(2pt)} &= f(\uparrow ,\downarrow )+f(\downarrow ,\uparrow )-\left(c^\dagger_{\downarrow ,1} c^\dagger_{\uparrow ,1} c_{\downarrow ,2} c_{\uparrow ,2}+c^\dagger_{\downarrow ,2} c^\dagger_{\uparrow ,2} c_{\downarrow ,1} c_{\uparrow ,1}\right)h_2 \\
		& + n_{\downarrow ,2} n_{\uparrow ,2}(2 h_1+h_2-2 h_4) + n_{\downarrow ,1} n_{\uparrow ,1}(2 h_1+h_2-2 h_6),\\
		\mathcal{H}_{12}^{(3pt)} & = n_{\uparrow ,1}  n_{\uparrow ,2}  \left( c^\dagger_{\downarrow ,1}  c_{\downarrow ,2}- c^\dagger_{\downarrow ,2}  c_{\downarrow ,1}\right)+n_{\downarrow ,1}  n_{\downarrow ,2}  \left( c^\dagger_{\uparrow ,1}  c_{\uparrow ,2}- c^\dagger_{\uparrow ,2}  c_{\uparrow ,1}\right),\\
	\end{aligned}
\end{equation}
with $ f $ defined
\begin{equation}\label{eq:4-osc}
	\begin{aligned}
		f(\alpha,\beta) &= \left(h_{10} \sigma -h_5 \tau \right) n_{\alpha,1} c^\dagger_{\beta,1} c_{\beta,2}-\left(h_3 \sigma +h_5 \tau \right) n_{\beta,2} c^\dagger_{\alpha,1} c_{\alpha,2}-\left(h_{10} \sigma +h_7 \tau \right) n_{\alpha,2} c^\dagger_{\beta,2} c_{\beta,1}\\
		&  + \left(h_3 \sigma -h_7 \tau \right) n_{\beta,1} c^\dagger_{\alpha,2} c_{\alpha,1}+h_9 c^\dagger_{\alpha,1} c^\dagger_{\beta,2} c_{\alpha,2} c_{\beta,1}+h_2 n_{\beta,2} n_{\alpha,1}.
	\end{aligned}
\end{equation}
We will focus on the case where $r_3\neq0$ (section \ref{r3no0}) since it contains the one-dimensional Hubbard model and the $S$-matrix of $AdS_5 \times S_5$. By using \eqref{condHam1} and \eqref{condHam2}, we should  diagonalise
\begin{align}
\mathcal{H}_{12}^{(osc)}=\mathcal{H}_{12}^{(1pt)}+\mathcal{H}_{12}^{(2pt)} - ((h_5+h_7)\tau+(h_{10}-h_3)\sigma)\mathcal{H}_{12}^{(3pt)}.
\end{align}
Similarly to \eqref{canonicaltransf}, we can write the canonical transformation
\begin{align}
&c_{\alpha,k}  = \frac{1}{\sqrt{{{}N}}} {{} \sum_{n=1}^N} e^{2\pi i \frac{k_\alpha n}{{{}N}} } \eta_{\alpha,n},
&&c_{\alpha,k}^\dag  = \frac{1}{\sqrt{{{}N}}} {{} \sum_{n=1}^N} e^{-2\pi i \frac{k_\alpha n}{{{}N}} } \eta_{\alpha,n}^\dag. \label{canonicaltransf16}
\end{align}
This is again the natural map since for periodic chains the one-particle eigenstates are simple plane waves. Then, rewriting densities $\mathbb{H}^{(k\,pt)}=\sum_{i=1}^{{}N} \mathcal{H}_{i,i+1}^{(k\,pt)},$ 
we arrive at
\begin{equation}
	\begin{aligned}
\mathbb{H}^{(1pt)}&=\left(h_1+h_2\right){{}N}^2+{{}N}\sum _{n=1}^{{}N} \left(h_4+h_6-2 \left(h_1+h_2\right)\right)N_n\\
&+\tau {{}N} \sum _{n=1}^{{}N} \left(\left(h_5+h_7\right) \cos \left(\frac{2 \pi  n}{{{}N}}\right)+i \left(h_5-h_7\right) \sin \left(\frac{2 \pi  n}{{{}N}}\right)\right)N_n,
	\end{aligned}
\end{equation}
where $N_n=\sum _{\alpha =\{\uparrow ,\downarrow \}}N_{\alpha,n}$, $N_{\alpha,n}=\eta^\dagger_{\alpha,n}\eta_{\alpha,n}$
and
\begin{equation}
	\begin{aligned}
\mathbb{H}^{(2pt)}&=\sum _{n,m=1}^{{}N}\left(2 h_2 \cos \left(\frac{2 \pi  (m+n)}{{{}N}}\right)+\left(2 \left(2 h_1+h_2\right)-2 \left(h_4+h_6\right)\right)\right) N_{\downarrow,n} N_{\uparrow,m}\\
&+\sum _{n,m=1}^{{}N}  4 h_2 \cos\left(\frac{2 \pi   (n-m)}{{{}N}}\right)N_{\downarrow,n}N_{\uparrow,m}\\
&-\sum _{n,m=1}^{{}N} 2  \left(\tau\left(h_5+h_7\right) \cos \left(\frac{2 \pi  m}{{{}N}}\right)+i \sin \left(\frac{2 \pi  m}{{{}N}}\right) \left(\left(h_3-h_{10}\right) \sigma +\left(h_5-h_7\right) \tau\right)\right)\\
&\left(N_{\downarrow,n}N_{\uparrow,m}+N_{\uparrow,n}N_{\downarrow,m}\right).
	\end{aligned}
\end{equation}
We are left with finding $\mathbb{H}^{(3pt)}$. After applying the canonical transformation \eqref{canonicaltransf16}, we were able to get a closed expression for ${{}N}=4,5$ and we get
\begin{equation}
	\begin{aligned}
\mathbb{H}^{(3pt)}&=i \sum_{\substack{\{\alpha,\beta\}=\{\uparrow,\downarrow\}, \\ \{\downarrow,\uparrow\}}}\Big[\sum _{n=1}^{{}N} N_{\alpha,n} \sin \left(\frac{2 \pi  n}{{{}N}}\right)\Big] \Big[\psi \sum _{i=1}^{{}N} \left(A_{i,i+1,i+{{}N}-1,i+2}^{(\beta)}+A_{i+2,i+{{}N}-1,i+1,i}^{(\beta)}\right)\\
&+\left(S_1 \sum _{i =1}^{{}N} N_{\beta,i } N_{\beta,i +1}+S_2 \sum _{i =1}^{{}N} N _{\beta,i } N _{\beta,i+2}\right)\Big],
	\end{aligned}
\end{equation}
where $A^{(\beta)}_{a,b,c,d}= \eta^\dagger_{\beta ,a}\cdot \eta^\dagger_{\beta ,b}\cdot \eta _{\beta ,c}\cdot \eta _{\beta ,d}$, $\psi$, $S_1$ and $S_2$ are constants dependent on ${{}N}$. We can see that $\mathbb{H}^{(3pt)}$ is not diagonal. 
The coefficient of $\mathcal{H}_{12}^{(3pt)}$ in \eqref{eq:ham1616} is $- ((h_5+h_7)\tau+(h_{10}-h_3)\sigma)$. Since $\mathbb{H}^{(3pt)}$ is not diagonal, if this coefficient is zero, our Hamiltonian will be explicitly of free fermion type. We evaluated it for {{} the} various models  of \cite{MCAAP2}, in particular for model 7 ($AdS_5\times  S^5$) and model 8 (which can be obtained from model 7 by taking a double limit). We found that for model 7, $((h_5+h_7)\tau+(h_{10}-h_3)\sigma)$ is not 0. For model 8, it can be easily shown that $((h_5+h_7)\tau+(h_{10}-h_3)\sigma)=k$, where $k$ is a constant. With a constant diagonal local basis transformation on the Hamiltonian, we can send $h_{10}\to {{} \zeta} h_{10}$ and $h_3\to \frac{h_3}{{{} \zeta}}$, with ${{} \zeta}$ constant, so the coefficient of the term $\mathbb{H}^{(3pt)}$ can be put to zero. In this case the Hamiltonian is diagonal and we obtained that a double limit of the Hamiltonian {{} $AdS_5\times  S^5$} is manifestly free fermion type.

Furthermore, we can notice that also the Hamiltonians of model 8 and 12 of \cite{MAAP} verify the conditions\footnote{We mention that \eqref{condHam1} and \eqref{condHam2} together with $((h_5+h_7)\tau+(h_{10}-h_3)\sigma)=0$ are the only conditions used to make the Hamiltonian \eqref{eq:ham1616}  free fermion type.} \eqref{condHam1} and \eqref{condHam2} and with a diagonal local basis transformation the coefficient of $\mathbb{H}^{(3pt)}$ can be set to zero. In these two cases the Hamiltonians are also of free fermion type. For model 12 this is not surprising since it corresponds to the free {{} Hubbard model ({\it i.e.} with only the kinetic term)} \cite{Korepin}.

\section{Conclusions}

In this paper we have shown how the new models classified in \cite{MCAAP} satisfy the free fermion condition. We collected some of the relevant literature which motivates how this condition  relates to the fact that all these models should be connected to theories of free fermions. We have then elucidated the power of this condition in reformulating the algebraic structure of lower-dimensional $AdS/CFT$ $R$-matrices in a novel and drastically simplified language of free fermion creation and annihilation operators, achieved by means of a suitable array of Bogoliubov transformations. Given that one of our examples is not nearest-neighbour, we conjecture that the free fermion condition must hold in an even more general setting.

We explicitly subdivided the models classified in \cite{MCAAP} into two classes (A and B), based on the single assignment of a constant, which is non-zero for class A satisfying the so-called Baxter condition, and zero for class B satisfying the free fermion condition. Such separation was reproduced by analysis and resolution of  the Sutherland equation. Moreover, we have shown how one can apply the free fermion condition to obtain a substitute to the algebraic Bethe ansatz in order to find the spectrum of 6-vertex class B models, and proceed instead with generalised inhomogeneous ansatz, from which one can also obtain homogeneous limit. We explicitly found the Bogoliubov transformation that always puts in all these cases the Hamiltonian in the manifest free fermion form, letting ourselves be guided by the intuition that comes from the traditional idea of the coordinate Bethe ansatz.

We have then applied these methods to the supersymmetric $AdS_3$ and $AdS_2$ cases in string theory, and found how they permit to rewrite known results in much more compact and suggestive forms. For the particular case of the 8-vertex B model, which is associated to the $AdS_2$ $R$-matrix, on two sites it was found what we call the {\textit{pseudo}-\textit{pseudo} vacuum}, by which we mean a vector which works just the same as the pseudo-vacuum (which such chains do not possess) and that exists only thanks to the particle-hole transformation which the free fermion condition has unveiled as possible.

As a next step, we studied $16 \times 16$ models, which contain the $AdS_5$ integrable model and the Hubbard chain. We showed how a free fermion type condition for such generic models arises. We then {{} tried} to recast the Hamiltonian in a form as close as possible to free fermions. One further direction is to investigate analogues of the free fermion condition for generalised $ AdS_{4,5} $ integrable models and their deformations, that could also potentially be useful for understanding the higher dimensional construction of Korepanov \cite{Korepanov_1993a,Korepanov_1993b,Zamolodchikov:1980} and its special limits. 

Our findings might shed some light for example on the result obtained in \cite{DiegoBogdanAle} for the central charge of the $CFT $ described by massless relativistic left-left and right-right $ R $-matrices of $AdS_3$ with pure RR flux. Such TBA was solved exactly and revealed a mysterious inner simplicity to the model, which might be related to its intimate free fermion nature. Likewise, the relative simplicity of the scalar products of Bethe states \cite{JuanMiguelAle} may naturally be related to the existence of a free fermion realisation\footnote{We thank Juan Miguel Nieto for communication on this point.}. Similarly, the results of \cite{Andrea2,Ale} are here {{} contextualised} within the free fermion general framework, and a new light is provided on the physical significance of their occurrence. 

\section*{Acknowledgements}

We thank Andrea Fontanella and Juan Miguel Nieto for discussions, comments and for careful reading of the manuscript. We thank Vladimir Korepin, Suvajit Majumdar, Rodrigo Pimenta, Alessandro Sfondrini, Bogdan Stefa\'nski and Konstantin Zarembo for useful discussions. We thank the anonymous referee for very useful comments. This work is supported by the EPSRC-SFI grant EP/S020888/1 {\it Solving Spins and Strings}. MdL, C.P., A.P. and ALR.  were furthermore supported by SFI and the Royal Society under grants UF160578, RGF$\backslash$EA$\backslash$181011 and RGF$\backslash$EA$\backslash$180167 and 18/EPSRC/3590. 

No data beyond those presented and cited in this work are needed to validate this study.  
\begin{appendix}

\section{Coordinate Bethe Ansatz for inhomogeneous chains}\label{app:BA}

The main advantage of the coordinate Bethe Ansatz is that it provides explicit wave functions, which are needed to map the system to free fermions. However, in order to deal with inhomogeneous spin-chains we need to use a slightly different version of the coordinate Bethe Ansatz that uses some of the properties of the algebraic Bethe Ansatz. In particular, it uses the $R$-matrix to derive the explicit form of the Bethe vectors. 

\paragraph{Formalism} The idea is to build explicit eigenvectors of the transfer matrix using a generalisation of a plane-wave type Ansatz. Consider a general transfer matrix build up out of an $R$-matrix $R(u,v)$

\begin{align}
T_N({ \theta_0},\vec{u}) = {\mbox{str}_0 \big[ R_{01} (\theta_0 , u_1) \ldots R_{0N} (\theta_0 , u_N) \big].}
\end{align}
We are interested in computing the explicit eigenvectors of $T_N$. For inhomogeneous spin-chains, all the commuting charges that $T_N$ generates are generically of range $N$ and hence the usual coordinate Bethe Ansatz approach does not work. In order to work around this, we will derive a different nearest neighbour condition that uses the $R$-matrix. 

Suppose we have an eigenvector $|v\rangle$ of the transfer matrix with eigenvalue $\Lambda$
\begin{align}
T _N\, |v\rangle = \Lambda\, |v\rangle.
\end{align}
By using the Yang-Baxter equation, it is easy to see that the $R$-matrix $R_{i,i+1}$ acts like a permutation operator on the transfer matrix\footnote{Here for simplicity we will omit the $_N$ on the transfer matrix $T$.}
\begin{align}
&R_{i,i+1}(u_i,u_{i+1}) T = T^{(i, i+1)} R_{i,i+1}(u_i,u_{i+1}),
&& T^{(i, i+1)}  := P_{i,i+1} T P_{i,i+1} \Big|_{u_i \leftrightarrow u_{i+1}}.
\end{align}
In other words, commuting the transfer matrix with $R_{i,i+1}$ results in permuting sites $i,i+1$ on the spin-chain. Now, since $\Lambda$ needs to be completely symmetric in the inhomogeneities $u_i$, we find that
\begin{align}
T^{(i, i+1)} \, R_{i,i+1}(u_i,u_{i+1})  \, |v\rangle = \Lambda^{(i, i+1)}\, R_{i,i+1}(u_i,u_{i+1})  \, |v\rangle.
\end{align}
In particular, $R_{i,i+1}(u_i,u_{i+1})  \, |v\rangle$  is an eigenvector of the \textit{permuted} transfer matrix.

But, on the other hand, the permuted eigenvector $|v^{(i, i+1)}\rangle$ is an eigenvector of the permuted transfer matrix by definition.  Thus we are naturally lead to the conditions that eigenvectors have the same transformation properties as the transfer matrix under the $R$-matrix and that $R_{i,i+1} |v\rangle $ and $|v^{(i, i+1)}\rangle$ need to be proportional
\begin{align}\label{eq:HybridEqn}
R_{i,i+1} |v\rangle = R_0 \,|v^{(i, i+1)}\rangle,
\end{align}
for some function $R_0$. By then assuming a factorised Bethe-type Ansatz for the wave function we see that \eqref{eq:HybridEqn} actually allows us to fix the Bethe vectors.

We will now work out this procedure explicitly for a general $R$-matrix of 6-vertex type \eqref{eq:Rosc}.

\paragraph{Vacuum} The ferromagnetic vacuum $|0\rangle$ trivially satisfies \eqref{eq:HybridEqn}, namely using the explicit form of the $R$-matrix \eqref{eq:Rosc} we find
\begin{align}
R_{i,i+1} |0\rangle = r_1(u_i,u_{i+1}) |0\rangle,
\end{align}
which shows that in this case $R_0 =  r_1(u_i,u_{i+1}) $. 

\paragraph{One magnon} The next step is to consider a state with a flipped spin, which takes the form
\begin{align}
|v\rangle = \sum_n A_n(v,\vec{u}) c^\dag_n|0\rangle.
\end{align}
This state has to satisfy \eqref{eq:HybridEqn} for $i=1,\ldots N$. Clearly, when the $R$-matrix does not act on the flipped spin, the state should behave like the vacuum. The easiest way to incorporate this is by making a factorised Ansatz of the form
\begin{align}\label{eq:Hybridmagnon}
A_n(v, \vec{u}) = f(v,u_n) \prod_{i=1}^{n-1} S(v,u_i).
\end{align}
This is an obvious generalisation of the usual coordinate Bethe Ansatz. Indeed setting all $u_i$ equal and taking $S= e^{i v}  $  reproduces the well-known plane-wave form of the Bethe vectors.

For concreteness, let us restrict to the first two sites and use the action of $R_{12}$ and denote $f_n(v) = f(v,u_n), S_n(v) = S(v,u_n)$. Plugging Ansatz \eqref{eq:Hybridmagnon} into the compatibility condition \eqref{eq:HybridEqn} then gives the following set of functional equations
\begin{align}
& f_1(v) r_3+ f_2(v) S_1(v) r_6  =  f_1(v) S_2 (v)r_1,
&&  f_2(v)S_1(v)r_2 + f_1(v) r_5 = f_2(v)r_1.
\end{align}
We have suppressed the dependence of $r_i$ on $u_{1,2}$. These give us a set of functional equations whose solution will depend on the explicit form of $r_i$. In particular, the solution can be read off from the Yang-Baxter equation to be
\begin{align}\label{eq:HybridfS}
&f_n(v) = \gamma \frac{r_6(u_n,v)}{r_2(u_n,v)},
&& S_n(v) =\frac{r_1(u_n,v)}{r_2(u_n,v)}.
\end{align}
The constant $\gamma$ simply determines the overall normalization of the vector and can be set to 1 without loss of generality. 

\paragraph{Multiple magnons} In the case of multiple magnons we need to introduce an $S$-matrix $\mathbb{S}$ that deals with the exchange of two magnons. Thus, inspired by the Bethe Ansatz, we make the following Ansatz
\begin{align}\label{eq:GenHybrid}
|v_1,v_2\rangle = \sum_{n<m} A_{n,m}(v_1,v_2,\vec{u}) c^\dag_n c^\dag_m |0\rangle,
\end{align}
where
\begin{align}
A_{n,m} = A_n(v_1) A_m(v_2) + \mathbb{S}(v_1,v_2) A_{n}(v_2) A_m(v_1).
\end{align}
We can again restrict to two sites and derive the following equation from \eqref{eq:HybridEqn} 
\begin{align}
-r_4\Big[f_1(v_1) f_2(v_2)S_1(v_2) + \mathbb{S}f_1(v_2) f_2(v_1)S_1(v_1) \Big] =- r_1\Big[f_2(v_1) f_1(v_2)S_2(v_2) + \mathbb{S}f_2(v_2) f_1(v_1)S_2(v_1) \Big],
\end{align}
where the $-$ sign on the right hand side comes from the fact that the permutation is graded. This gives a simple linear equation for $\mathbb{S}$ which is easy to solve
\begin{align}
\mathbb{S} = -\frac{r_1(v_1,v_2) r_4(v_1,v_2)}{r_2(v_1,v_2)  r_3(v_1,v_2) - r_5(v_1,v_2) r_6(v_1,v_2)}.
\end{align}
We see upon using the free fermion condition this simply reduces to $-1$ as expected.

\paragraph{Bethe equations}
Finally, imposing periodicity leads to the following Bethe equations
\begin{align}
\label{auxillo}
\prod_{n=1}^N S(v_i,u_n) = - \prod_{m=1}^M \mathbb{S}(v_i,v_m).
\end{align}
In the homogeneous limit this reduces to the normal Bethe equations. If we interpret the inhomogeneities as particles with momenta, then we can further impose
\begin{align}
e^{ip_iN} = \prod_{n=1}^M S(u_i,v_n)
\end{align}
and the level matching condition
\begin{align}
\prod_{i=1}^N e^{ip_i} =1.
\end{align}
However, the last two conditions are not needed for the diagonalization of the transfer matrix.

\paragraph{Covectors}

Following the exact same reasoning as above we can also determine the covectors of the transfer matrix. These are needed in the canonical transformation to free fermions. Thus, we consider the dual vacuum $\langle 0 |$ and excited states of the form $\langle 0 | c_i$ so that
\begin{align}
1 = \langle 0|0\rangle = \langle 0 | c_i c^\dag_i |0\rangle = \ldots.
\end{align}
For covectors we should use $R^{-1}$ instead of $R$ in our considerations. By using braiding unitarity, we see that
\begin{align}
\langle 0 | R^{-1}_{12}(u_1,u_2) = \langle 0 | \frac{1}{r_1}
\end{align}
and hence  we find $R_0 = 1/r_1$. Let us denote the coefficients of the Bethe covectors by $f^*, S^*$ and $\mathbb{S}^*$ then an analogous computation shows that
\begin{align}
&f^*_n(v) = \gamma^* \frac{r_5(u_n,v)}{r_1(u_n,v)},
&& S^*_n(v) =\frac{r_2(u_n,v)}{r_1(u_n,v)} = \frac{1}{S_n(v)},
&&\mathbb{S}^* = \mathbb{S}^{-1}.
\end{align}
As expected, we see that the Bethe equations are the same for vectors and covectors.

{{}
\section{$AdS_3$ transfer matrix $T_2$ from the general formula\label{appB}}
In this appendix we specialise the general formula {{} (\ref{inhocano})} to the massless pure Ramond-Ramond $AdS_3$ case in the particular case of $T_2$. The two-site formula reduces to
\begin{eqnarray}
\label{expre}
&&c_1 = \frac{f_1^*(v_1) S_1(v_1) S_2(v_1)}{\sqrt{|f_1(v_1)|^2 +|f_2(v_1)|^2}} \eta_1 + \frac{f_1^*(v_2) S_1(v_2) S_2(v_2)}{\sqrt{|f_1(v_2)|^2 +|f_2(v_2)|^2}} \eta_2,\nonumber\\
&&c_2 = \frac{f_2^*(v_1) S_2(v_1)}{\sqrt{|f_1(v_1)|^2 +|f_2(v_1)|^2}} \eta_1 + \frac{f_2^*(v_2) S_2(v_2)}{\sqrt{|f_1(v_2)|^2 +|f_2(v_2)|^2}} \eta_2,
\end{eqnarray}
where, taking into the account the fermionic signs, the specific case of the $R$-matrix we are dealing with implies
\begin{eqnarray}
f_i(v) = \frac{r_6(\theta_i,v)}{r_2(\theta_i,v)} = - \mbox{csch} \frac{\theta_i - v}{2}, \qquad S_i(v) = \frac{r_1(\theta_i,v)}{r_2(\theta_i,v)}=\mbox{coth}\frac{v - \theta_i}{2}. 
\end{eqnarray}
The $v_i$s are the solutions to the auxiliary Bethe equations (\ref{auxo}), which coincide with (\ref{aux}) upon {{} taking the reciprocal of both sides of the equation}:
\begin{eqnarray}
1 = \prod_{i=1}^2 \mbox{tanh} \frac{v_n-\theta_i}{2}, \qquad n = 1,2, 
\end{eqnarray}
namely
\begin{eqnarray}
\label{acco}
v_1 = -\infty, \qquad v_2 = + \infty.
\end{eqnarray} 
By taking the limits of the expressions (\ref{expre}) according to (\ref{acco}) we obtain
\begin{eqnarray}
&&c_1 = -\frac{e^{\frac{\theta_2}{2}}}{\sqrt{e^{\theta_1} + e^{\theta_2}}} \eta_1 +\frac{e^{\frac{\theta_1}{2}}}{\sqrt{e^{\theta_1} + e^{\theta_2}}} \eta_2,\nonumber\\
&&c_2 = \frac{e^{\frac{\theta_1}{2}}}{\sqrt{e^{\theta_1} + e^{\theta_2}}} \eta_1 +\frac{e^{\frac{\theta_2}{2}}}{\sqrt{e^{\theta_1} + e^{\theta_2}}} \eta_2, 
\end{eqnarray}
implying
\begin{eqnarray}
c_1 \propto \eta_1 - e^{\frac{\theta_1-\theta_2}{2}} \eta_2, \qquad 
c_2 \propto \eta_1 + e^{\frac{\theta_2-\theta_1}{2}} \eta_2.
\end{eqnarray}  
By considering that $\cot \alpha = - e^{\frac{\theta_2 - \theta_1}{2}}$, it is easy to see that this is canonically equivalent to the transformation (\ref{tend}) upon redefining $\eta_2 \to - \eta_2$ (and consequently for $\eta_1^\dagger$ and $\eta_2^\dagger$).
}

\section{Recursive formulas for the $AdS_3$ transfer matrix\label{appC}}
\label{appendB}
In this appendix, we shall produce explicit recursive formulas - by-passing the intermediate step of the auxiliary Bethe equations - for the transfer matrix with a generic number of sites, exploiting the (though deceptively) simple form {{} (\ref{surprise}) of the massless pure Ramond-Ramond $AdS_3$} $R$-matrix
\begin{eqnarray}
\label{therea}
R_{0i}(\theta_0 - \theta_i) = \cosh \frac{\theta_{0i}}{2} \Big[1-2 \eta_{0i}^\dagger \eta_{0i}\Big] \equiv \cosh \frac{\theta_{0i}}{2} \Big[1-2 N_{0i}\Big].
\end{eqnarray}
In (\ref{therea}) we have renamed the operators $\eta_1$ and $\eta^\dagger_1$ in  (\ref{surprise}) as $\eta_{0i}$ and $\eta^\dagger_{0i}$, respectively, to display the fact {{} that they are the operators $\eta_1(2)$ and $\eta_2(2)$ with inhomogeneities adapted to the isolated pair of sites $0$ and $i$. This shows how the complication of the $R$-matrix is in fact hidden in the definition of these operators}\footnote{Let us point out that these operators differ from the $\eta_i$ and $\eta^\dagger_i$ operators introduced in the main text, which should be the result of a global transformation involving all sites of the chain of particles, and should directly diagonalise the transfer matrix.}.

The transfer matrix is therefore simply given by
\begin{eqnarray}
\label{assemble}
&&T_N = \Bigg[\prod_{i=1}^N \cosh \frac{\theta_{0i}}{2} \Bigg] \mbox{str}_0 \Big[1 - 2 N_{01}\Big]... \Big[1 - 2 N_{0N}\Big] = \\
&& \Bigg[\prod_{i=1}^N \cosh \frac{\theta_{0i}}{2}\Bigg] \Bigg(\langle 0_0| \Big[1 - 2 N_{01}\Big]... \Big[1 - 2 N_{0N}\Big] |0_0\rangle - \langle 0_0| c_0 \Big[1 - 2 N_{01}\Big]... \Big[1 - 2 N_{0N}\Big] c^\dagger_0|0_0\rangle \Bigg),\nonumber
\label{T_N}
\end{eqnarray}
where we have denoted by $|0_0\rangle$ the state $|\phi\rangle$ in the auxiliary space $0$. It is clear that one needs to find a way to compute the generic expressions 
\begin{eqnarray}
\langle 0_0 | N_{01} ... N_{0m}|0_0\rangle \quad \mbox{and} \quad \langle 0_0 | c_0 N_{01} ... N_{0m} c^\dagger_0 |0_0\rangle
\end{eqnarray}
for generic $m = 1,..,N$ to be able to calculate $T_N$. 

For these quantities we are able to find a combined system of recursive relations. Let us begin with the first one, and let us define two objects:
\begin{eqnarray}
X_m \equiv \langle 0_0 | N_{01} ... N_{0m}|0_0\rangle, \qquad Y_m \equiv \langle 0_0 | N_{01} ... N_{0m} c_0^\dagger|0_0\rangle.
\end{eqnarray}
One can easily see that the following holds true:
\begin{eqnarray}
&&\langle 0_0 | N_{01} ... N_{0m}|0_0\rangle = \langle 0_0 | N_{01} ... N_{0,m-1} \eta^\dagger_m \eta_m |0_0\rangle = \langle 0_0 | N_{01} ... N_{0,m-1}\eta^\dagger_m (\alpha_m c_0 + \beta_m c_m)|0_0\rangle = \nonumber\\
&&\qquad \qquad \beta_m \langle 0_0 | N_{01} ... N_{0,m-1}(\alpha_m c_0^\dagger + \beta_m c_m^\dagger)|0_0\rangle c_m=\nonumber\\
&&\qquad \qquad \qquad \qquad \alpha_m \beta_m \langle 0_0 | N_{01} ... N_{0,m-1}c_0^\dagger|0_0\rangle c_m + \beta_m^2 \langle 0_0 | N_{01} ... N_{0,m-1}|0_0\rangle n_m,\nonumber
\end{eqnarray} 
where we have used the fact that $|0_0\rangle$ is a bosonic state, and we have defined
\begin{eqnarray}\label{abth}
\alpha_i \equiv \cos \alpha_{0i}, \qquad \beta_i \equiv \sin \alpha_{0i}, \qquad \cot 2 \alpha_{0i} = \sinh \frac{\theta_{0i}}{2}.
\end{eqnarray}
This results in the following recursive relation
\begin{eqnarray}
\label{X-rec}
X_m = \alpha_m \beta_m Y_{m-1} c_m + \beta_m^2 X_{m-1} n_m
\end{eqnarray}
Similar manipulations performed starting from $\langle 0_0 | N_{01} ... N_{0m} c_0^\dagger|0_0\rangle$ bring to another relation:
\begin{eqnarray}
Y_m = \alpha_m^2 Y_{m-1} + \alpha_m \beta_m X_{m-1} c^\dagger_m + \beta_m^2 Y_{m-1} n_m
\label{Y-rec}
\end{eqnarray}
Given the starting points\footnote{The recursion could actually be set to start from $m=0$ with initial values $X_0 =1$ and $Y_0=0$, and it would automatically generate (\ref{start}).}
\begin{eqnarray}
\label{start}
X_1 = \beta_1^2 n_1, \qquad Y_1 = \alpha_1 \beta_1 c_1^\dagger,
\end{eqnarray}
with (\ref{X-rec}) and (\ref{Y-rec}) one generates all the terms in (\ref{T_N}). The next few terms are for instance straightforwardly obtained:
\begin{eqnarray}
X_2 = \beta_1^2 \beta_2^2 n_1 n_2 + \alpha_1 \beta_1 \alpha_2 \beta_2 c_1^\dagger c_2, \qquad Y_2 = \alpha_1 \beta_1 c_1^\dagger (\alpha_2^2 + \beta_2^2 n_2) + \beta_1^2 \alpha_2 \beta_2 n_1 c_2^\dagger.
\end{eqnarray}
By resolving the recursion system (\ref{X-rec})-(\ref{Y-rec}) in terms of suitable operators and performing a reindexing one can obtain two closed separated equations for $ X_m $ and $ Y_m $, respectively, which results in
\begin{equation}\label{X-Y}
	\begin{aligned}
		X_m =& \prod _{i=0}^{m-1} \Omega _{2,i+1} + \sum _{i=0}^{m-1} \Omega _{1,i+1} \left(\prod _{j=i}^{m-2} \Omega _{2,j+2}\right) \sum _{\mu =0}^{i-1}  \Omega _{3,\mu +1} \left(\prod _{\nu =\mu }^{i-2} \left(\Omega _{4,\nu +2}+\Omega_{2,\nu +2}\right)\right) X_{\mu }, \\
		Y_m =& \sum _{i=0}^{m-1} \Omega _{3,i+1} \left(\prod _{j=i}^{m-2} \left(\Omega _{4,j+2}+\Omega _{2,j+2}\right)\right) \left(\prod _{\alpha =0}^{i-1} \Omega _{2,\alpha +1}+\sum _{\alpha =0}^{i-1}  \Omega _{1,\alpha +1} \left(\prod _{\beta =\alpha }^{i-2} \Omega _{2,\beta +2}\right)	Y_\alpha\right) ,\\
		& \text{for } m \geq 0, \qquad \Omega _{1,q} = \alpha _q \beta _q c_q, \quad \Omega _{2,q} = \beta _q^2 n_q, \quad\Omega _{3,q}=\alpha _q \beta _q c_q{}^{\dagger }, \quad\Omega _{4,q}=\alpha _q^2,
	\end{aligned}
\end{equation}
where $ \alpha_i $, $ \beta_i $ are defined in (\ref{abth}). Notice that all the terms on the right hand side of (\ref{X-Y}) have to be understood as being normal-ordered $ :\cdot: $, {\it i.e.} the fermionic operators $ c_q $ and $ c^\dagger_q $ with lower $ q $ appear to the left. Alternatively, by eliminating $ Y_m $ using (\ref{Y-rec}) as
\begin{eqnarray}
Y_m = \sum_{i=1}^m \alpha_i \beta_i \, X_{i-1} \, c_i^\dagger \prod_{j=i+1}^m (\alpha_j^2 + \beta_j^2 n_j), \qquad m \geq 1, \qquad Y_0 = 0,
\end{eqnarray}
and substituting back into (\ref{X-rec}), we obtain a recursion only for $X_m$ (which are the objects ultimately relevant to the transfer matrix):
\begin{eqnarray}
X_m = \beta_m^2 X_{m-1} n_m + \alpha_m \beta_m \Bigg[\sum_{i=1}^{m-1} \alpha_i \beta_i \, X_{i-1} \, c_i^\dagger \prod_{j=i+1}^{m-1} (\alpha_j^2 + \beta_j^2 n_j)\Bigg] c_m, \qquad m \geq 1, \qquad X_0=1.
\end{eqnarray}
By defining
\begin{eqnarray}
\nu_m \equiv \beta_m^2 n_m, \quad \psi_i^\dagger \equiv \alpha_i \beta_i c_i^\dagger, \quad \xi_{im} \equiv \Bigg[\prod_{j=i+1}^{m-1} (\alpha_j^2 + \beta_j^2 n_j) \Bigg]\alpha_m \beta_m c_m
\end{eqnarray}
we can rewrite the recursion more compactly as
\begin{eqnarray}
X_m = x_{m-1} \nu_m + \sum_{i=1}^{m-1} X_{i-1} \psi^\dagger_i \xi_{im},\qquad m \geq 1, \qquad X_0=1. 
\end{eqnarray}
Experimenting with up to $m=5$ leads us to conjecture that the general solution is obtained by sprinkling {\it block-molecules} $\Gamma_{ij} = \psi^\dagger_i \xi_{ij}$ in a sea of $\nu$'s. For instance,
\begin{eqnarray}
&&X_5 = \nu_1 \nu_2 \nu_3 \nu_4 \nu_5 + \Gamma_{12} \nu_3 \nu_4 \nu_5 + \nu_1 \Gamma_{23} \nu_4 \nu_5 + \Gamma_{13} \nu_4 \nu_5 + \Gamma_{14} \nu_5 + \nu_1 \Gamma_{24} \nu_5  + \nu_1 \nu_2 \Gamma_{34}\nu_5 \\
&& + \Gamma_{12} \Gamma_{34} \nu_5 + \Gamma_{15} + \nu_1 \Gamma_{25} + \nu_1 \nu_2 \Gamma_{35} + \Gamma_{12} \Gamma_{35} + \nu_1 \nu_2 \nu_3 \Gamma_{45} + \Gamma_{12} \nu_3 \Gamma_{45} + \Gamma_{13}\Gamma_{45} + \nu_1 \Gamma_{23} \Gamma_{45}\nonumber.
\end{eqnarray}   
It is also easy to see that, by further defining
\begin{eqnarray}
\widehat{X}_m \equiv (-2)^m X_m, \qquad \widehat{\Gamma}_{ij} \equiv (-2)^{j-1+1} \Gamma_{ij}, \qquad \widehat{\nu}_j \equiv -2 \nu_j,  
\end{eqnarray}
we have that $\widehat{X}_m$ - which is what ultimately enters the expansion (\ref{assemble}) - is obtained by sprinkling block-molecules $\widehat{\Gamma}_{ij}$ in a sea of $\widehat{\nu}$'s. This way we have conveniently reabsorbed all the factor of $-2$ and the expansion (\ref{assemble}) will have all coefficients equal to $1$.

With analogous reasoning we can obtain the recursive relations for the other part of the supertrace. We simply state the result: by defining
\begin{eqnarray}
Z_m \equiv \langle 0_0 | c_0 N_{01} ... N_{0m} c_0^\dagger|0_0\rangle, \qquad {{} L_m} \equiv \langle 0_0 |c_0 N_{01} ... N_{0m} c_0 c_0^\dagger|0_0\rangle,
\end{eqnarray} 
we can derive
\begin{eqnarray}
Z_m = \alpha_m^2 Z_{m-1} + \alpha_m \beta_m {{} L_{m-1}} c_m^\dagger + \beta_m^2 Z_{m-1}n_m, \qquad {{} L_m} = \alpha_m \beta_m Z_{m-1}c_m + \beta_m^2 {{} L_{m-1}} n_m,
\label{UZ-rec}
\end{eqnarray}
and deriving closed separated recursions in analogy with (\ref{X-Y}) we obtain
\begin{equation}\label{U-Z}
	\begin{aligned}
		{{} L_m} =& \sum _{i=0}^{m-1} \Xi_{4,i+1} \prod _{j=i}^{m-2} \Xi_{3,j+2} \left(\prod _{\alpha =0}^{i-1} \left(\Xi_{1,\alpha
			+1}+\Xi_{3,\alpha +1}\right)+\sum _{\beta =1}^{i-1} \Xi_{2,\beta +1} \prod _{\alpha =\beta }^{i-2} \left(\Xi_{1,\alpha +2}+\Xi_{3,\alpha +2}\right) U_{\beta } \right)\\
		Z_m =& \prod _{i=0}^{m-1} \left(\Xi_{1,i+1}+\Xi_{3,i+1}\right) + \sum _{j=1}^{m-1} \Xi_{2,j+1} \left(\prod _{i=j}^{m-2} \left(\Xi_{1,i+2}+\Xi_{3,i+2}\right)\right) \sum _{\mu =0}^{j-1} \Xi_{4,\mu +1} \left(\prod _{\nu =\mu }^{j-2} \Xi_{3,\nu +2}\right) Z_{\mu},\\
		&  \text{for } m\geq 0 \text{, } \quad \Xi _{1,s}=\alpha _s^2, \quad \Xi _{2,s}=\alpha _s \beta _s c_s{}^{\dagger }, \quad \Xi _{3,s} = \beta _s^2 n_s, \quad \Xi_{4,s} = \alpha _s \beta _s c_s.
	\end{aligned}
\end{equation}
Alternatively we can again eliminate ${{} L_m}$ from the second formula in (\ref{UZ-rec}), to get
\begin{eqnarray}
{{} L_m} = \sum_{i=1}^m \alpha_i \beta_i \, Z_{i-1} \, c_i \prod_{j=i+1}^m \beta_j^2 n_j, \qquad m \geq 1, \qquad U_0 = 0.
\end{eqnarray}
By again substituting into the first formula in (\ref{UZ-rec}), we obtain a recursion only for $Z_m$ which is what matters for the transfer matrix:
\begin{eqnarray}
Z_m = Z_{m-1} (\alpha_m^2 + \beta_m^2 n_m) + \alpha_m \beta_m \Bigg[\sum_{i=1}^{m-1} \alpha_i \beta_i \, Z_{i-1} \, c_i \prod_{j=i+1}^{m-1} \beta_j^2 n_j \Bigg]c^\dagger_m, \qquad m \geq 1, \qquad Z_0=1.
\end{eqnarray}
We can therefore see that by defining
\begin{eqnarray}
\mu_m \equiv \alpha_m^2 + \beta_m^2 n_m, \quad \zeta_i^\dagger \equiv \alpha_i \beta_i c_i, \quad \rho_{im}^\dagger \equiv \Bigg[\prod_{j=i+1}^{m-1} \beta_j^2 n_j \Bigg]\alpha_m \beta_m c_m^\dagger
\end{eqnarray}
we can rewrite this recursion more compactly as well in the form
\begin{eqnarray}
Z_m = x_{m-1} \mu_m + \sum_{i=1}^{m-1} Z_{i-1} \zeta_i \rho^\dagger_{im},\qquad m \geq 1, \qquad Z_0=1. 
\end{eqnarray}
This is virtually the same recursion as for the $X_m$ {\it mutatis mutandis}, hence the solution is again conjectured to be obtained by sprinkling block-molecules $\Psi_{ij} \equiv \zeta_i \rho^\dagger_{ij}$ is a sea of $\mu$'s in much the same way.
Likewise, by further defining
\begin{eqnarray}
\widehat{Z}_m \equiv (-2)^m Z_m, \qquad \widehat{\Psi}_{ij} \equiv (-2)^{j-1+1} \Psi_{ij}, \qquad \widehat{\mu}_j \equiv -2 \mu_j,  
\end{eqnarray}
we have that $\widehat{Z}_m$ - which enters the expansion (\ref{assemble}) - is obtained by sprinkling block-molecules $\widehat{\Psi}_{ij}$ in a sea of $\widehat{\mu}$'s.

This appears as a very efficient way of generating the transfer matrix at arbitrary $N$. Of course one still needs to assemble these components into the sort of binomial expansion coming from (\ref{assemble}), which is not a trivial task. In the end, already at $N=3$ we are faced with a rather complicated expression nevertheless. We have first checked that at $N=2$ we reproduce the result (\ref{unruly}) in the main text via
\begin{eqnarray}
&&T_2 = \cosh \frac{\theta_{01}}{2} \cosh \frac{\theta_{02}}{2} \, \Big[1 - 2 X_1 - 2 \tilde{X}_1 + 4 X_2 - (1 - 2 Z_1 - 2 \tilde{Z}_1 + 4 Z_2) \Big] = \nonumber\\
&&= \cosh \frac{\theta_{01}}{2} \cosh \frac{\theta_{02}}{2} \, \Big[1 + \widehat{X}_1 + \tilde{\widehat{X}}_1 + \widehat{X}_2 - (1 + \widehat{Z}_1 + \tilde{\widehat{Z}}_1 + \widehat{Z}_2) \Big]
\end{eqnarray} 
where we have denoted with $\tilde{X}_1$ the quantity $\langle 0_0 | N_{02}|0_0\rangle$, which can be obtained with no difficulties by appropriate replacements once $X_1$ is known (likewise for the hatted quantities). 
Similarly is done with $\tilde{Z}_1$. 
Then, we find a rather bulky formula for $T_3$:
\begin{eqnarray}
&&T_3 \propto 2 \sum_{i=1}^3 \alpha_i^2 + 4 \Big[\alpha_1 \beta_1 \alpha_2 \beta_2\Big(c_1^\dagger c_2 (1 - 2 \beta_3^2 n_3) - c_1 c_2^\dagger  (1 - 2 \alpha_3^2 - 2 \beta_3^2 n_3)\Big)\nonumber\\
&&\qquad \qquad \quad \qquad + \alpha_1 \beta_1 \alpha_3 \beta_3\Big(c_1^\dagger c_3 (1 - 2 \alpha_2^2 - 2 \beta_2^2 n_2) - c_1 c_3^\dagger  (1 - 2 \beta_2^2 n_2)\Big)\nonumber\\
&&\qquad \qquad \qquad \quad + \alpha_2 \beta_2 \alpha_3 \beta_3\Big(c_2^\dagger c_3 (1 - 2 \beta_1^2 n_1) - c_2 c_3^\dagger  (1 - 2 \alpha_1^2 - 2 \beta_1^2 n_1)\Big)\Big]\nonumber\\
&&-8\Big[\beta_1^2 \beta_2^2 \beta_3^2 n_1 n_2 n_3 - (\alpha_1^2 + \beta_1^2 n_1)(\alpha_2^2 + \beta_2^2 n_2)(\alpha_3^2 + \beta_3^2 n_3)\Big]\nonumber\\
&&- 4\Big[(\alpha_1^2 + \beta_1^2 n_1)(\alpha_2^2 + \beta_2^2 n_2) + (\alpha_1^2 + \beta_1^2 n_1)(\alpha_3^2 + \beta_3^2 n_3) +(\alpha_2^2 + \beta_2^2 n_2)(\alpha_3^2 + \beta_3^2 n_3)\Big].
\end{eqnarray}
Ultimately, readjusting terms as described in the case of $\tilde{X}_1$ above becomes progressively more and more cumbersome as $N$ increases.

\section{Examples of Hamiltonians}
\label{examples}
In this {{} appendix}, we provide two examples from $AdS_3$ which exemplify the general procedure outlined in {{} section \ref{omogenee}}.

\subsection{Pure Ramond-Ramond case}

\subsubsection{Open chain\label{rro}}

In this short section we show how to diagonalise the Hamiltonian with free fermion operators for an open chain. Although for $AdS_3$ purposes this is not directly applicable, it is interesting to note that in the pure Ramond-Ramond case the Hamiltonian reads \cite{Andrea}
\begin{eqnarray}
{{}\mathbb{H}}_N = \sum_{n=1}^{N-1} h_n, \qquad h_n = i(c_n^\dagger c_{n+1} + c_n c_{n+1}^\dagger).  
\end{eqnarray}
We have verified this result on two sites using the new variables $(\eta_i,\eta^\dagger_i)$, where one needs to pay particular attention to the $\theta$-dependence of the new operators themselves. In particular, one immediately notices that the transformation which diagonalises $T_2$ is not suitable to diagonalise ${{}\mathbb{H}}_2$. This can be seen in the new variables, since taking the derivative w.r.t. $\theta$ breaks up $N_1$ into something which ceases to be diagonal in the new basis.

Nevertheless, the Hamiltonian is Hermitian and can be diagonalised, one simply needs a different unitary transformation. It is easy to see that the map\footnote{{}We shall not use the symbol $\eta$ for the open chain, since the general theory of section \ref{analternativetotheABA} was developed for closed chains.}
\begin{eqnarray}
\psi_\pm = \frac{1}{\sqrt{2}} (c_1 \pm i c_2), \qquad \psi^\dagger_\pm = \frac{1}{\sqrt{2}} (c_1^\dagger \mp i c_2^\dagger), 
\end{eqnarray}
transforms the Hamiltonian on two sites to
\begin{eqnarray}
{{}\mathbb{H}}_2 = \psi^\dagger_+ \psi_+ -  \psi^\dagger_- \psi_-.
\end{eqnarray}
This map can be easily checked to be canonical, namely the new operators still satisfy canonical anticommutation relations.
 
We also remark that the Hamiltonian ${{}\mathbb{H}}_2$ here differs from the one of \cite{Lieb}, which is more akin to $c_1^\dagger c_2 - c_1 c^\dagger_2$ (still being Hermitian). The transformation suggested in that paper will not work here, even though we believe that there must surely be some sort of telescopic transformation on all $N$ sites which achieves a global free fermion diagonal form in our case as well. For instance, we can go one step further and diagonalise ${{}\mathbb{H}}_3$, which equals
\begin{eqnarray}
{{}\mathbb{H}}_3 = i (c_1^\dagger c_2 + c_1 c_2^\dagger + c_2^\dagger c_3 + c_2 c_3^\dagger).
\end{eqnarray}
The following unitary transformation
\begin{eqnarray}
\psi_- = \frac{1}{2}(- c_1 - i \sqrt{2} \, c_2 + c_3), \qquad \psi_+ = \frac{1}{2}(- c_1 + i \sqrt{2} \, c_2 + c_3), \qquad \psi_3 = \frac{1}{\sqrt{2}} (c_1 + c_3), 
\end{eqnarray}
achieves the free fermion expression
\begin{eqnarray}
{{}\mathbb{H}}_3 = \sqrt{2} (\psi^\dagger_- \psi_- -  \psi^\dagger_+ \psi_+). 
\end{eqnarray}
Any unitary map $\psi_i = U_i^j c_j$ is clearly canonical\footnote{{} The symbol $\psi_N$ should not be confused with the same symbol appearing in the main text in the overall factor of the transfer matrix. We trust that the context will be sufficient to discriminate between the two.}: $\{\psi_i,\psi^\dagger_j\} = U_j^m \big[U^n_j\big]^* \{c_m,c^\dagger_n\} = U_j^m \big[U^\dagger\big]_m^j = \delta^j_i$.

The structure of the 1-particle sub-block of the Hamiltonian is always the following: 
\begin{eqnarray}
{{}\mathbb{H}}_{N}^{{}(1pt)} = \begin{pmatrix}0&1&0&0&0&0&....\\ -1&0&1&0&0&0&....\\0&-1&0&1&0&0&....\\0&0&-1&0&1&0&....\\...\end{pmatrix}.
\end{eqnarray}
The diagonalisation of this sub-block always appears to produce the result
\begin{eqnarray}
\label{half}
{{}\mathbb{H}}_{N}^{{}(1pt)} = \sum_{n=1}^N \, \Big[2 \cos \frac{\pi n }{N+1}\Big] \, \psi_n^\dagger \psi_n = \sum_{n=1}^N \, \Big[2 \sin \frac{q_n}{2}\Big] \, \psi_n^\dagger \psi_n,
\end{eqnarray}
where we have defined 
\begin{eqnarray}
q_n \equiv \pi \frac{N+1 - 2n}{N+1}.
\end{eqnarray} 
  
We conjecture that (\ref{half}) is in fact the complete Hamiltonian:
\begin{eqnarray}
\label{half2}
{{}\mathbb{H}}_N= \sum_{n=1}^N \, \Big[2 \cos \frac{\pi n }{N+1}\Big] \, \psi_n^\dagger \psi_n = \sum_{n=1}^N \, \Big[2 \sin \frac{q_n}{2}\Big] \, \psi_n^\dagger \psi_n.
\end{eqnarray}
We can make a connection with gapless spin-chains and spinon-excitations as noticed in \cite{Andrea} - here with quantised momenta on a finite chain, see also \cite{Lieb}. The form of the free fermion operators $\psi_n$ grows complicated with $N$, and is determined using the explicit form of the eigenvectors. The cases $N=2$ and $N=3$ fit into this framework, with an appropriate name-redefinition of the generators. We have verified the one-particle eigenvalue pattern up to $N=7$, and the form (\ref{half2}) explicitly for the $N=4$ case as well with the help of Mathematica\footnote{In the case of $N=4$, the unitary transformation reads
\begin{eqnarray}
\begin{pmatrix}\psi_1\\\psi_2\\\psi_3\\\psi_4\end{pmatrix}=\begin{pmatrix}\frac{1}{\sqrt{5+\sqrt{5}}}&\frac{i(-1-\sqrt{5})}{2\sqrt{5+\sqrt{5}}}&\frac{-1-\sqrt{5}}{2\sqrt{5+\sqrt{5}}}&\frac{i}{\sqrt{5+\sqrt{5}}}\\\frac{-1}{\sqrt{5-\sqrt{5}}}&\frac{i(-1+\sqrt{5})}{2\sqrt{5-\sqrt{5}}}&\frac{1-\sqrt{5}}{2\sqrt{5-\sqrt{5}}}&\frac{i}{\sqrt{5-\sqrt{5}}}\\\frac{1}{\sqrt{5-\sqrt{5}}}&\frac{i(-1+\sqrt{5})}{2\sqrt{5-\sqrt{5}}}&\frac{-1+\sqrt{5}}{2\sqrt{5-\sqrt{5}}}&\frac{i}{\sqrt{5-\sqrt{5}}}\\\frac{-1}{\sqrt{5+\sqrt{5}}}&\frac{i(-1-\sqrt{5})}{2\sqrt{5+\sqrt{5}}}&\frac{1+\sqrt{5}}{2\sqrt{5+\sqrt{5}}}&\frac{i}{\sqrt{5+\sqrt{5}}}\\\end{pmatrix}\begin{pmatrix}c_1\\c_2\\c_3\\c_4\end{pmatrix}.
\end{eqnarray}}. 

Notice that roughly half of the states have negative energy according to (\ref{half2}) - for instance, for even $N$ this occurs for $n = \frac{N}{2}+1 ,...,N$. For these states, we need to perform a particle-hole transformation in the standard fashion to have all particles and antiparticles with positive massless dispersion relation
\begin{eqnarray}
\epsilon(q) = 2 \, \Big\vert \sin \frac{q}{2}\Big\vert.
\end{eqnarray}


\subsubsection{Closed chain\label{rr}}

It is easy to provide an example of the closed-chain treatment performed in the text in the $AdS_3$ case as well. The closed chain has two extra terms corresponding to ${{}\mathbb{H}}_{N}^{{}(1pt)}$, and the appropriate transformation is displayed in {{} \ref{canonicaltransf}}. For instance, for three sites
\begin{eqnarray}
{{}\mathbb{H}}_3 = i(c_1^\dagger c_2 + c_1 c_2^\dagger + c_2^\dagger c_3 + c_2 c_3^\dagger + c_3^\dagger c_1 + c_3 c_1^\dagger).
\end{eqnarray}
The unitary map for general $N$ is given {{} (after an inconsequential relabelling of the $\eta_i$)} by 
\begin{eqnarray}
\label{mappag}
c_k^\dagger = \frac{1}{\sqrt{N}} \sum_{n=1}^N e^{\frac{2 \pi i}{N}k(n-1)} \eta^\dagger_n, \qquad c_k = \frac{1}{\sqrt{N}} \sum_{n=1}^N e^{-\frac{2 \pi i}{N}k(n-1)} \eta_n. 
\end{eqnarray}
We can see how this map works explicitly for $N=3$. By plugging the expansion in the Hamiltonian we obtain
\begin{eqnarray}
{{}\mathbb{H}}_3 = \frac{i}{3} \sum_{k,m,n=1}^3 \Bigg[e^{\frac{2\pi i}{3}[k(n-1) - (k+1)(m-1)]}\eta^\dagger_n\eta_m + e^{-\frac{2\pi i}{3}[k(n-1) - (k+1)(m-1)]}\eta_n\eta^\dagger_m\Bigg].
\end{eqnarray} 
We can perform the sum over $k$ first, obtaining
\begin{eqnarray}
\sum_{k=1}^3 \, {e^{\frac{2\pi i}{3}[k(n-1) - (k+1)(m-1)]}}_{\Big\vert 3+1 \equiv 1} = 3 e^{\frac{2\pi i}{3}(1-m)} \delta_{n,m},
\end{eqnarray}
therefore
\begin{eqnarray}
{{}\mathbb{H}}_3 = i \sum_{m=1}^3 \Bigg[e^{\frac{2\pi i}{3}(1-m)}\eta^\dagger_m\eta_m + e^{-\frac{2\pi i}{3}(1-m)}\eta_m\eta^\dagger_m \Bigg]= 2 \sum_{m=1}^3 \Big[\sin \frac{2 \pi}{3}(m-1) \Big] \eta^\dagger_m\eta_m.
\end{eqnarray} 
We have used the commutation relations of the fermionic operators to combine the two terms into one, with the contribution from the $1$ in the commutator vanishing due to a complete sum of roots of unity.
 
The result can be generalised to arbitrary $N$ based on the properties of the roots of unity as exploited in {{} section \ref{omogenee}}, revealing again a spectrum similar to the one of the open chain:
\begin{eqnarray}
{{}\mathbb{H}}_N = 2 \sum_{m=1}^N \Big[\sin \frac{2 \pi}{N}(m-1) \Big] \eta^\dagger_m\eta_m.
\end{eqnarray} 
We have explicitly tested these particular formulas up to $N=4$.

\subsection{Mixed Flux case}

\subsubsection{Open chain\label{rrmo}}

The open-chain mixed flux case seems to work as well, thanks to the fact that $R(0)$ equals the graded permutation in this case too, however the complication of the functional form of the $R$-matrix forces us to stop much earlier in $N$. Let us just treat the two cases which we can manage with Mathematica at the moment.

For $N=2$ the Hamiltonian, obtained with the same method as in the previous section (and with a convenient normalisation), reads
\begin{eqnarray}
{{}\mathbb{H}}_2 = -i \Big[R^{-1}(\theta)  \frac{d}{d\theta}R(\theta)\Big]_{\vert \theta=0} = [\cos 2 \omega]\, (n_1 + n_2) + c_1 c_2^\dagger - c_1^\dagger c_2, \qquad n_i = c_i^\dagger c_i, \qquad i=1,2,
\end{eqnarray}
where we have defined
\begin{eqnarray}
\omega \equiv \frac{\pi}{2k}.
\end{eqnarray}
We remind that the external parameter $k=2,3,...$, is a fixed natural number which measures the mixture of the fluxes in the string theory\footnote{The case $k=1$ is peculiar, see \cite{gamma2} for details.}.
The Hamiltonian ${{}\mathbb{H}}_2$ is Hermitian and more similar to the one studied in \cite{Lieb}, except for the $\cos 2\omega$ term. 

Via the Bogoliubov transformation
\begin{eqnarray}
\psi_\pm = \frac{1}{\sqrt{2}}(c_1 \pm c_2)
\end{eqnarray}
we easily get
\begin{eqnarray}
{{}\mathbb{H}}_2 = 2 \Big([\cos^2 \omega] \,\psi^\dagger_- \psi_- - [\sin^2 \omega] \,\psi^\dagger_+ \psi_+\Big).
\end{eqnarray}

The $N=3$ case reads
\begin{eqnarray}
{{}\mathbb{H}}_3 = [\cos 2\omega] (n_1 + 2 n_2 + n_3) + c_1 c_2^\dagger - c_1^\dagger c_2 + c_2 c_3^\dagger - c_2^\dagger c_3.
\end{eqnarray}
In order to get some feeling of how the Hamiltonian work we can show the matrix in the one-particle sector:
\begin{eqnarray}
{{}\mathbb{H}}_{3}^{{}(1pt)} = \begin{pmatrix}\cos 2\omega&-1&0\\-1&2 \cos 2\omega&-1\\0&-1&\cos 2\omega\end{pmatrix}.
\end{eqnarray}
In fact, in general the one-particle sector will look like this:
\begin{eqnarray}
{{}\mathbb{H}}_{N}^{{}(1pt)} = \begin{pmatrix}\cos 2\omega&-1&0&0&0&0&....\\ -1&2 \cos 2\omega&-1&0&0&0&....\\0&-1&2 \cos 2\omega&-1&0&0&....\\0&0&-1&2 \cos 2\omega&-1&0&....\\...\\0&0&0&0&0&-1&\cos 2\omega\end{pmatrix}.
\end{eqnarray}
In the case of $N=3$ we can show with the help of the computer that the orthogonal transformation 
\begin{eqnarray}
&&\psi_3 = \frac{c_3 - c_1}{\sqrt{2}}, \qquad \psi_- = \frac{4(c_1 + c_3) - 2 \, c_2 \cos 2\omega + \sqrt{2} \, c_2 \sqrt{17 + \cos 4\omega}}{2\sqrt{17 + \cos 4\omega - \sqrt{2} \cos 2\omega \sqrt{17 + \cos 4\omega}}},\nonumber\\
&&\psi_+ = \frac{4(c_1 + c_3) - 2 \, c_2 \cos 2\omega - \sqrt{2} \, c_2 \sqrt{17 + \cos 4\omega}}{2\sqrt{17 + \cos 4\omega + \sqrt{2} \cos 2\omega \sqrt{17 + \cos 4\omega}}}
\end{eqnarray}
exactly transforms the Hamiltonian into
\begin{eqnarray}
{{}\mathbb{H}}_3 = [\cos 2\omega] \psi_3^\dagger \psi_3 + \Big[\frac{1}{4}(6 \cos 2\omega - \sqrt{2}\sqrt{17 + \cos 4\omega}) \Big] \psi^\dagger_- \psi_- + \Big[\frac{1}{4}(6 \cos 2\omega + \sqrt{2}\sqrt{17 + \cos 4\omega})\Big] \psi^\dagger_+ \psi_+.\nonumber
\end{eqnarray}
Proceeding with general $k$ appears to be rather involved. Only the case $k=2$ (which is closely related to the properly-regularised $k=1$ case \cite{gamma2}) simplifies drastically. In fact, with a suitable normalisation, the full mixed-flux Hamiltonian ${{}\mathbb{H}}_N$ at $k=2$ reduces to a special case of the one diagonalised in \cite{Lieb} (it corresponds to setting the parameter $\gamma$ of \cite{Lieb} to $0$, and mapping $N$ here to $N-1$ in \cite{Lieb}).

\paragraph{General $N$} Solving it for general $N$ goes as follows. First, since the Hamiltonian is quadratic in the fermionic oscillators $c$ and only contains terms of the form $c^\dag c$, it is enough to restrict to the one-particle sector. Diagonalizing the Hamiltonian then gives a linear map between the standard basis vectors $c^\dag_i |0\rangle$ and the eigenvectors which we will denote by $\eta^\dag_i |0\rangle$. The same holds true for the covectors $\langle 0| c_i$ and  $\langle 0| \eta_i$. By construction, the Hamiltonian then takes the form
\begin{align}
{{}\mathbb{H}}_N = 2\sum_{i=1}^N (\cos2\omega -\cos \tau_i)  \eta^\dag_i \eta_i,
\end{align}
where $\tau_i$ correspond to the solutions of
\begin{align}
\label{cheby}
P_\tau = \frac{\sin (N+1) \tau }{\sin \tau }  - 2 \cos 2
   \omega  \frac{ \sin N \tau }{\sin \tau } +  \cos^2 2
   \omega  \frac{\sin (N-1) \tau }{\sin \tau }=0.
\end{align}
It is not hard to see that this can be rewritten as a polynomial equation in $\cos\tau$ which has exactly $N$ solutions. In fact, by definition of the Chebyschev polynomials of the second kind
\begin{eqnarray}
\label{byuse}
U_m(\cos \tau) \equiv \frac{\sin (m+1)\tau}{\sin \tau},
\end{eqnarray}
we see that (\ref{cheby}) identifies the zeroes of a combination of Chebyschev polyomials with highest degree $N$:
\begin{eqnarray}
P_\tau = U_N(\cos \tau)  - 2 \cos 2
   \omega  \, U_{N-1}(\cos \tau) +  \cos^2 2
   \omega  \, U_{N-2}(\cos \tau)=0.
\end{eqnarray}
Using the recursion between different $U_m$'s we may also write the above as
\begin{eqnarray}
P_\tau = 2 (\cos \tau - \cos 2 \omega) U_{N-1}(\cos \tau) - \sin^2 2 \omega \, U_{N-2}(\cos \tau) =0.
\end{eqnarray} 
The case $k=2$ of course drastically simplifies, and one has the very compact
\begin{eqnarray}
U_N(\cos \tau)=0,
\end{eqnarray}
singling out the zeroes of a single Chebyschev polyomials of degree $N$. By using (\ref{byuse}), such zeroes can be chosen in correspondence of the values  
\begin{eqnarray}
\tau_n = \frac{n \pi}{N+1}, \qquad n = 1,...,N, 
\end{eqnarray}
which produces the Hamiltonian \cite{Lieb}
\begin{eqnarray}
{{}\mathbb{H}}_N = - 2 \sum_{n=1}^N \Big[\cos \frac{n \pi}{N+1} \Big] \, \eta_n^\dagger \eta_n. 
\end{eqnarray}
The minus sign which implies negative energy for half the states can again be dealt with by a standard particle-hole transformation \cite{Lieb}.

\subsubsection{Closed chain}

In the case of the closed chain the problem simplifies drastically and one is reduced to the general theory we have developed in {{} section \ref{omogenee}}, since the Hamiltonian simply becomes
\begin{eqnarray}
{{}\mathbb{H}}_N = 2 [\cos 2 \omega] \sum_{i=1}^N n_i + c_1 c_2^\dagger - c_1^\dagger c_2 + c_2 c_3^\dagger - c_2^\dagger c_3  ... +  c_{N-1}c_N^\dagger - c_{N-1}^\dagger c_N + c_1 c_N^\dagger - c_1^\dagger c_N. \nonumber    
\end{eqnarray}
The very first term is proportional to the identity and simply shifts the energies of $2 \cos 2 \omega$. The remaining non-diagonal part is easily diagonalised if we simply notice that it is very similar to the pure Ramond-Ramond case of section \ref{rr}, except for a relative minus sign between the two main terms (and no multiplying factor of $i$). Namely, if we take for instance the $N=3$ case, we now have, with the very same eigenvectors, that
\begin{eqnarray}
{{}\mathbb{H}}_3 = 2 [\cos 2 \omega] \mathfrak{1} - \sum_{m=1}^3 \Bigg[e^{\frac{2\pi i}{3}(1-m)}\eta^\dagger_m\eta_m - e^{-\frac{2\pi i}{3}(1-m)}\eta_m\eta^\dagger_m \Bigg]= 2 [\cos 2 \omega] \mathfrak{1}- 2 \sum_{m=1}^3 \Big[\cos \frac{2 \pi}{3}(m-1) \Big] \eta^\dagger_m\eta_m.\nonumber
\end{eqnarray} 
One can then straightforwardly generalise to 
\begin{eqnarray}
{{}\mathbb{H}}_N = 2 [\cos 2 \omega] \mathfrak{1}- 2 \sum_{m=1}^N \Big[\cos \frac{2 \pi}{N}(m-1) \Big] \eta^\dagger_m\eta_m = - 2 \sum_{m=1}^3 \Big[\cos \Big(\frac{2 \pi}{3}(m-1)\Big) - \cos 2 \omega\Big] \eta^\dagger_m\eta_m, 
\end{eqnarray}
with the exact same map (\ref{mappag}) still holding. 

\end{appendix}

\end{document}